\documentclass[11pt]{article}
\usepackage{mymacros}
\usepackage{caption}
\newcommand{\coin}{\textsc{coin}}
\newcommand{\inst}{\textsc{instr}}

\newcommand{\CalX}{\mathcal{X}}
\renewcommand{\phi}{\varphi}
\newcommand{\good}{\textsc{good}}
\newcommand{\goodone}{\textsc{index}}
\newcommand{\close}{\textsc{close}}

\newcommand{\gam}[1]{\Gamma_{#1}(k_1,k_2)}
\newcommand{\thmax}{\widetilde{H}_{\max}}
\newcommand{\phmax}{{\widehat{H}_{\max}}}
\newcommand{\wtt}[1]{\widetilde{#1}}

\title{{\bf Centralised multi link measurement compression with
side information}}

\author{
Sayantan Chakraborty${}^\ddag$
\and
Arun Padakandla${}^*$
\thanks{
${}^*$ /
Department of Electrical Engineering and Computer Science,
University of Tennessee at Knoxville, U.S.A.
Email: {\sf arunpr@utk.edu}.
}
\and
Pranab Sen${}^\ddag$
\thanks{
${}^\ddag$ 
School of Technology and Computer Science, 
Tata Institute of Fundamental Research, Mumbai, India.
Email: {\sf \{kingsbandz,pranab.sen.73\}@gmail.com}.
Sayantan is partially supported by a Google India late Ph.D. fellowship.
Sayantan and Pranab are
supported by the Department of Atomic Energy, 
Government  of India, under project no. RTI4001.
}
 }

\date{}

\begin{document}
\maketitle

\begin{abstract}
We prove new one shot achievability results for measurement compression
of quantum instruments with side information at 
the receiver. Unlike previous one shot results for this problem, our
one shot bounds are nearly optimal and
do not need catalytic randomness. In fact, we state a more
general problem called centralised multi link measurement compression
with quantum side information
and provide one shot achievability results for it. As a simple corollary,
we obtain one shot measurement compression results for quantum
instruments with side information that we  mentioned earlier.
All our one shot results lead to the standard results for this
problem in the asymptotic iid setting. We prove our achievability
bounds by first proving
a novel sequential classical quantum multipartite covering lemma,
which should be of independent interest.
\end{abstract}

\section{Introduction}\label{sec:introduction}
In order to obtain important statistical information about quantum systems, an experimenter must perform \emph{measurements} on the system. This information can be used in subsequent physical operations on the system. However, the measurement process is inherently noisy. This noise could arise due to \emph{extrinsic} noise in the measurement procedure itself (which is uncorrelated to the state) and/or \emph{intrinsic} noise arising from the uncertainty introduced by the quantum state being measured. An important information theoretic problem is to filter out the extrinsic noise so that we can compress the number of bits required to describe the measurement outcome.

We model the measurement procedure using the positive operator valued measure (POVM) formalism. Consider the following setting: The experimenter Alice possesses a quantum state $\rho$ and a POVM $\Lambda$. The action of the POVM on the quantum state is to produce a classical register (in Alice's possession) which contains an index corresponding to the measurement outcome, as well as a post measurement quantum state (to which Alice does not have access). The task is for Alice to send the contents of her classical register to Bob, such that the correlations between this register and the post measurement state are conserved, using as few bits as possible. This problem is formally known as \emph{measurement compression}.

In his seminal paper, Winter \cite{Winter} observed that the above task can be achieved if one can construct an approximate decomposition of the POVM of the following form :
\[
\Lambda\overset{\eps}{\approx} \sum_c P_{\coin}(c)\Lambda(c)
\]
where each $\Lambda(c)$ is a POVM with fewer possible classical outcomes as compared to $\Lambda$. The variable $c$ corresponds to the outcome of a coin toss which is distributed according to $P_{\coin}$. As long as Alice and Bob both have access to the coin $c$, Alice only needs to send enough bits to be able to describe the set of possible outcomes of $\Lambda(c)$. Note that this distribution is independent of the state $\rho$, and thus is a source of extrinsic noise. The measurement procedure can then be summarised as follows :
\begin{enumerate}
    \item Toss a coin $\sim P_{\coin}$. Let the outcome be $c$.
    \item Measure $\rho$ with $\Lambda(c)$
\end{enumerate}

Let the number of bits needed to describe the state of the coin be $C$ and the number of bits of communication from Alice to Bob be $R$. Winter characterised the region of all $(R,C)$ pairs for which the above decomposition exists. A cleaner exposition of Winter's results can be found in a later paper by Wilde et al. \cite{Wilde}, who also extend Winter's original contribution by considering the case when Bob has access to some part of the quantum state being measured. This protocol is known as Measurement Compression with Quantum Side Information or MC-QSI in short. In this scenario Alice and Bob share the quantum state $\rho^{AB}$, where Alice possesses the system $A$ and bob possesses the system $B$. Alice then measures her share of the state and sends her classical register to Bob. Wilde et al. showed that in this setting, Bob can use his share of the quantum state to reduce the number of bits Alice needs to send to him even further.

All of the above works consider the measurement compression problem in the asymptotic iid regime i.e. in the limit of a large  number of repeated identical measurements on several copies of the original quantum state. A more general variant of the problem would be if only one copy of the state was available for measurement. This is the so called \emph{one-shot} regime. In a recent paper, Anshu, Jain and Warsi \cite{AnshuJain} provide an achievable region for MC-QSI in this regime. In contrast to all previous works, that paper shows a reduction of the MC-QSI problem to the well studied problem of \emph{quantum message compression with side information} \cite{ConvexSplitLemma}, which allows them to leverage known tools for the latter problem. However, even though Anshu et al.'s results imply the best 
known asymptotic iid bounds for measurement compression, their one-shot 
bounds are unsatisfactory, due to the following issues:
\begin{enumerate}

\item 
The protocol uses shared randomness in a catalytic manner i.e. 
it needs some a large amount of shared randomness to begin with 
but regenerates part
of this randomness at the end of the protocol. In the asymptotic iid
limit with $n$ copies, this protocol is applied iteratively $n/m$
times to blocks of states containing $m$ copies each, where $m < n$.
By choosing $m$ appropriately, the authors could show
that that the amount of extra initial randomness required is $o(n)$. 
Thus, the amount of extra initial shared randomness per iteration goes to
zero in the asymptotic iid limit. Nevertheless in the one shot scenario 
the extra initial shared randomness can potentially  be very large, 
almost as
large as the alphabet size;

\item 
The second and perhaps more serious issue is that the bounds 
achieved by Anshu et al. are optimal only in the case when the 
probability distribution induced by the original POVM is uniform 
on the alphabet $X$. In all other cases, there is a trade-off between 
the amount of shared randomness required  to run the protocol and 
the extent to which Alice can compress her message to Bob.

\end{enumerate}

In this paper we define an even
more general problem called {\em centralised multi-link 
measurement compression
with side information} and prove a one shot achievability result for it.
The problem that we discussed above will be called the {\em point to
point} message compression problem.
As corollaries of our result for centralised multi link measurement 
compression, we obtain
\begin{enumerate}

\item
New one shot achievability result for point to point measurement 
compression with
side information in both feedback as well as non-feedback cases;

\item 
Our protocol does not require any extra initial shared randomness to 
begin with;

\item 
It recovers the optimal bounds of Wilde et al. in the asymptotic 
iid limit;

\item 
It achieves the natural one-shot bounds that one expects for this 
problem, for any probability distribution on the set of outcomes, without 
making any compromises between the amount of shared randomness required 
and the rate of compression. 

\end{enumerate}

To motivate this problem we consider a natural generalisation of the measurement compression, to the case of quantum instruments. Instruments are the most general model for quantum measurements, which include both a classical output and a post measurement quantum state \cite{inst1,inst2}. A quantum instrument is a CPTP map $\mathcal{N}_{\inst}$ of the following form:
\[
\mathcal{N}_{\inst}(\rho)\coloneqq \sum_x\ketbra{x}^X\otimes \mathcal{N}_x(\rho) 
\]
where each $\mathcal{N}_x$ is a completely positive trace non-increasing map with Kraus decompositions as follows 
\begin{align*}
\mathcal{N}_x(\rho)&= \sum_y \mathcal{N}_{x,y}~\rho~\mathcal{N}_{x,y}^{\dagger} \\
\intertext{and}
\sum_y\mathcal{N}_{x,y}^{\dagger}~\rho~\mathcal{N}_{x,y}&\leq \I
\end{align*}
 One can then ask the question  whether we can prove a compression theorem for quantum instruments as well. This theorem was claimed without proof in \cite{DevetakHarrowWinter} and proven rigorously in \cite{Wilde}. Wilde et al. solve this problem by suitably reducing it to an instance of the original measurement compression problem. Their approach at simulating the action of $\mathcal{N}_{\inst}$ is to identify its operation as a tracing out of the $Y-$register of the state
\[
\sum_{x,y}\ketbra{x}^X\otimes \ketbra{y}^Y\otimes\mathcal{N}_{x,y}~\rho~\mathcal{N}_{x,y}^{\dagger}.
\]
Naturally, they design a POVM simulation protocol wherein Alice simulates the POVM $\{\mathcal{N}_{x,y}^{\dagger}\mathcal{N}_{x,y}\}_{x,y}$. Recognizing that Bob only needs to recover the $X-$register they propose Alice discards her simulation of the $Y-$register and thereby achieve rates that one would naturally expect. 

\begin{figure}[hhh]
    \centering
    \includegraphics[scale=.7]{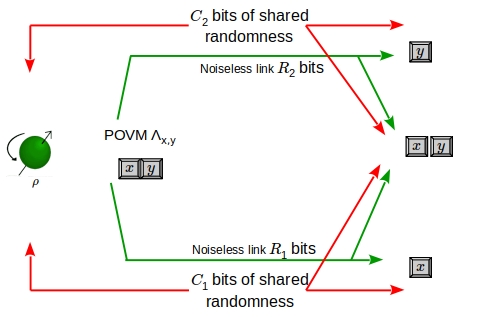}
    \caption{Fig.1}
    \label{fig:centralised}
\end{figure}

We now define our new problem called the
centralised multilink measurement compression problem with side
information. Taking a cue from the discussion above we consider a POVM $\Lambda_{XY}=\brak{\Lambda_{x,y}}$ which outputs two classical symbols $x$ and $y$ according to some joint distribution which depends on the state $\rho$. There exist two separate noiseless channels, called $X$ and $Y$ channels,
from Alice to Bob, and 
two independent public coin registers, called $X$ and $Y$ public coins,
between them. We use the word \emph{link} to refer to a noiseless channel.
During the protocol at most one of the links may be turned OFF by 
an adversary without Alice or Bob's knowledge. We ask whether we can design a \emph{single} simulation protocol that enables Bob to recover either $X$ or $Y$ or both $X,Y$ depending on which link is ON. See Fig ~\ref{fig:centralised}. 

To be precise, we seek \textit{one} simulation protocol that enables Bob to recover (i) the $X$ register when only the $X$-link of rate $R_{1}$ is active and they share $C_{1}$ bits of randomness, (ii) the $Y-$register when only the $Y$-link of rate $R_{2}$ is active and they share $C_{2}$ bits of randomness, and (iii) recover both $X$ and $Y$ registers when both links are active and they share a total of $C_{1}+C_{2}$ random bits. We require that Alice's and Bob's strategies should be agnostic to
which links are operational i.e. their encoding and decoding strategies
should continue to work even if one link fails.

A careful consideration of this problem reveals the following challenge. The requirement that each link enable recovery of the corresponding classical register precludes the use of designing conditional codebooks for the POVM simulation protocol. Otherwise, if the codebook for $Y$ were to be conditionally dependant on the codebook for $X$, Bob would be unable to recover $Y$ when the link of rate $R_1$ is inactive, and vice-versa. Therefore, this strategy, which works well for the purposes of Wilde et al. , fails in our case.

On the other hand, if the codebooks have to be designed independently, then it is unclear how Bob would be able to recover both $X,Y$ when both links are active. We overcome this issue by proving a novel quantum covering lemma, referred to herein as \emph{measure transformed sequential covering lemma}. This lemma is one of the main technical contributions of our work. The issue in proving a covering lemma of this sort is that most arguments invariably run into the issue of \emph{simultaneous smoothing}, which is an outstanding open problem in quantum information theory \cite{Drescher}. Recently however, Chakraborty, Nema and Sen \cite{ChakrabortyNemaSen} showed how one can get around the smoothing problem in certain situations. We adapt their construction and generalise it for our purposes. In fact, the proof of the covering lemma presented in this work is much easier and far more general than that in \cite{ChakrabortyNemaSen}, and it may be useful in other applications as well.

Before stating our one shot achievability results
for centralised measurement compression, we need to recall the
concept of rate splitting in the quantum setting 
\cite{ratesplitting}.
The joint probability distribution
on the measurement outcomes $(x,y)$ of the given POVM $\Lambda$
is defined by
$
p(x,y) := 
\Tr[((\Lambda_{x,y} \otimes \I^{BR})(\rho))^{A B R}],
$
which in turn defines $p(x)$ and $p(y|x)$ in the natural fashion.
Now, the splitting function creates two new random variables $U$ and $V$
supported on alphabet $\CalX$, defines
$x := \max\{u,v\}$, and splits the probability distribution $p(x)$
into a joint distribution $p^\theta(u,v)$ parametrised by a real number
$\theta \in [0,1]$. This splitting function
satisfies the following properties for any $\theta \in [0,1]$:

\begin{enumerate}
\item $U^\theta$, $V^\theta$ are independent random variables;
\item The probability distribution of the random variable 
$\max\{U^\theta,V^\theta\}$ is the same as that of the random variable $X$;
\item When $\theta = 0$, $U^0 = X$ and $V^0$ is constant. When $\theta = 1$, $V^1 = X$ and $U^1$ is constant.
\end{enumerate}

We prove the following theorem for one shot achievability results
for centralised measurement compression. The theorem is stated in terms
of smooth one shot entropic quantities like
$
\phmax^{\eps}(\cdot),
$
$
I_{\max}^{\eps}(\cdot : \cdot),
$
$
I_{H}^{\eps}(\cdot : \cdot),
$
the definitions of which can be found in the full version of this paper. They are the natural one shot analogues
of Shannon entropic quantities like entropy and mutual information.
\begin{theorem}{\bf (Centralised multi link measurement compression)}
\label{thm:CentralisedMeasurementCompression}
Consider the $\epsilon$-error centralised multi-link measurement 
compression problem with side
information at Bob in the feedback case.
One achievable rate region is obtained as the
union over a parameter $\theta \in [0,1]$ of the regions 
$S_\theta$ defined by:
\[
S_\theta~:~ 
\begin{aligned}
&R_X  ~~=~~  R_U + R_V \\
&\begin{aligned}R_U ~~ > ~~  &I_{\max}^{\eps}(U:RB) - I^\epsilon_H(U:B)  \\
	  &+ O(\log \epsilon^{-1})\end{aligned} \\
&\begin{aligned}R_Y  ~~>~~  &I_{\max}^{\eps}(Y:RBU)  - I^\epsilon_H(Y:B) \\
	  &+ O(\log \epsilon^{-1})\end{aligned} \\
&\begin{aligned}R_V  ~~>~~  &I_{\max}^{\eps}(V:RBUY) - I^\epsilon_H(V:B) \\
	  &+ O(\log \epsilon^{-1})\end{aligned} \\
&C_X  ~~=~~  C_U + C_V \\
C_U + &R_U  ~~>~~  \phmax^{\eps}(U)-I_H^{\eps}(U:B) \\
C_Y + &R_Y  ~~>~~  \phmax^{\eps}(Y)-I_H^{\eps}(Y:B) \\
C_V + &R_V  ~~>~~  \phmax^{\eps}(V)-I_H^{\eps}(V:B), \\
\end{aligned}
\]
where the entropic quantities are calculated for the control state
\[
\begin{aligned}
\sum\limits_{(u,v,y)\in \CalX \times \CalX \times \mathcal{Y}}
&p^\theta(u) p^\theta(v) p(y|u,v)
\ketbra{u,v,y}^{UVY} \\
&\otimes 
\frac{((\Lambda_{\max\{u,v\},y} \otimes \I^{BR})(\rho))^{A B R}}
     {\Tr[((\Lambda_{\max\{u,v\},y} \otimes \I^{BR})(\rho))^{A B R}]}.
     \end{aligned}
\]
The above state is 
obtained by {\em splitting} random variable $X$ into independent random 
variables $U$, $V$ in the state
$
\sum\limits_{(x,y)\in \CalX \times \mathcal{Y}}
\ketbra{x,y}^{XY} \otimes ((\Lambda_{x,y} \otimes \I^{BR})(\rho))^{A B R}
$
according to the parameter $\theta$. 
Another achievability
region is obtained by rate splitting $Y$ instead of $X$. The total
achievable region is the union of the two regions. The encoding
and decoding strategies are agnostic to which links are actually
functioning.
\end{theorem}
Our general theorem above lends itself to two important corollaries
which are stated below.
\begin{corollary}\label{cor:iidrateRegion}
In the asymptotic iid setting of the centralised measurement compression
problem, the following rate region per channel
use is achievable:
\begin{eqnarray*}
\label{eq:rateRegion}
R_{X} &>& I(X : BR) - I(X:B)  \\
R_{Y} &>& I(Y : BR) - I(Y:B)  \\
R_{X}+R_{Y} &>& I(XY : BR) + I(X:Y) \\
& &- I(X:B) - I(Y:B)\\
R_{X}+C_{X} &>& H(X) - I(X:B) \\
R_{Y}+C_{Y} &>& H(Y) - I(Y:B), 
\end{eqnarray*}
where all entropic quantities are computed with 
respect to 
$
\sum\limits_{(x,y)\in \CalX \times \mathcal{Y}}
\ketbra{x,y}^{XY} \otimes ((\Lambda_{x,y} \otimes \I^{BR})(\rho))^{A B R}.
$
\end{corollary}

\begin{corollary}\label{cor:sideinfo}
For the point to point measurement compression problem with
side information for quantum
instruments , an achievable region can be obtained by assuming
that the Y links is not functional. In the one shot setting we get
\begin{align*}
R_X &> I_{\max}^{\eps}(X:RB) - I^\epsilon_H(X:B) 
	  + O(\log \epsilon^{-1}) \\
\intertext{and}
R_{X}+C_{X} &> \phmax^\eps(X) - I^\epsilon_H(X:B).
\end{align*}
In the asymptotic iid setting this reduces to
\begin{align*}
R_X > I(X:RB) - I(X:B) 
	  + O(\log \epsilon^{-1}) 
\intertext{and}
R_{X}+C_{X} > H(X) - I(X:B).
\end{align*}
Above, all entropic quantities are computed with 
respect to 
$
\sum\limits_{(x,y)\in \CalX \times \mathcal{Y}}
\ketbra{x,y}^{XY} \otimes ((\Lambda_{x,y} \otimes \I^{BR})(\rho))^{A B R}.
$
\end{corollary}

 \subsection{Measurement Compression with Quantum Side Information}\label{sec:MCQSI}
 
 In this section we provide a brief exposition of the ideas which allow us to prove \cref{cor:sideinfo}. The techniques developed in this section allow us to design a measurement compression protocol in which the receiver Bob can use quantum side information to reduce the umber of bits that Alice needs to send him. The setup is as follows : \\
 Alice and Bob share a joint quantum state $\rho^{AB}$, where Alice has access to system $A$ and Bob has access to system $B$. Alice is given a POVM $\Lambda^A=\brak{\Lambda_x^A}$, where each POVM element $\Lambda_x$ acts on the system $A$ and corresponds to a classical outcome $x\in\CalX$. We define the \emph{true} post measurement state
 \begin{align*}
     \sigma^{RBX\hat{X}}&\coloneqq \Tr_A\left[\left(\I^{RB}\otimes \Lambda^A\right)\phi^{RAB}\right] \\
     &\begin{aligned}= \sum\limits_{x\in \CalX}&\ketbra{x,x}^{X\hat{X}}\otimes \\ &\Tr_A\left[\left(\I^{RB}\otimes \Lambda^A_x\right)\phi^{RAB}\right]
     \end{aligned}
 \end{align*}
We allow Alice and Bob access to a noiseless forward classical channel and also to public coins. The system $\hat{X}$ is used by Bob to store his guess for the outcome of the measurement. In the case of the true post measurement state, one can imagine that Alice sent Bob the full description of the outcome through the classical channel, which Bob then stored in $\hat{X}$. Now suppose that 
Alice and Bob use a protocol $\mathcal{P}$ which allows them to reduce the number of bits which need to be communicated to Bob. Let the post measurement state created by $\mathcal{P}$ be $\Tilde{\sigma}^{RBX\hat{X}}$. We require that for any such protocol,
\[
\norm{\sigma-\Tilde{\sigma}}_1 \leq \eps
\]

The first idea is that Alice designs a class of POVMs $\brak{\theta_k}$, where each POVM $\theta_k$ is indexed by the setting of the public coin $k\in [2^C]$. Each $\theta_k$ contains fewer outcomes than the original $\Lambda$ which allows Alice to save on the number of random bits that she needs to describe the outcome. In particular, each $\theta_k$ acts on the system $A$ and produces as output an index in the set $[2^{R+B}]$ i.e. 
\[
\theta_k\coloneqq \brak{\theta_k(\ell)~|~\ell\in [2^{R+B}]}
\]
. Winter's original construction of these POVMs relied on the Operator Chernoff \cite{AhlswedeWinter} bound to show the existence of these POVMs. Unfortunately, the Operator Chernoff bound is not easily adapted to the one shot setting. Instead we use a new one shot covering lemma (\cref{subsec:coveringLemma}) along with a one shot operator inequality (\cref{subsec:opineq}) to show the existence of the compressed POVM. These two techniques are new and are powerful enough to allow us to emulate the Operator Chernoff bound even in the one shot setting. We believe that these two new tools maybe useful elsewhere. We note that recently Padakandla \cite{padakandla} provided another alternate way of designing the compressed POVM in the one shot setting.

The second idea is that after Alice applies a POVM $\theta_k$ and observes some outcome $\ell$, she hashes the outcome into a bit string by using a hash function $f : [2^{R+B}]\to [2^R]$, which is chosen uniformly at random from a $2$-universal hash family. This hash value is what Alice sends to Bob. Suppose that the hash value sent by Alice is $m$.

Upon receiving Alice's bit string, Bob chooses a POVM $\gamma(m,k)$, based on Alice's message and the setting of the public coin. Bob will use this POVM to measure his system $B$ to find the index which is consistent with the outcome $\ell$. The idea is that as long as the set $f^{-1}(m)$ is not too large, Bob should be able to find the correct measurement outcome with a high probability of success.

The above protocol is essentially a classical data compression with quantum side information protocol (CDC-QSI), which has been studied in the asymptotic iid setting by Devetak and Winter \cite{DevetakWinter} and Renes \cite{RenesIID} and in the one shot setting by Renes and Renner \cite{RenesRenner}. However, it is not immediately obvious how one can adapt Renes and Rener's protocol to our setting. Instead, we design a new one shot protocol for CDC-QSI, which is more suitable for our setting. The one shot rates our new protocol achieves are different from those of Renes and Renner, but recover to the same asymptotic iid bounds achieved by them. The upshot of our protocol is that it is \emph{composable} with the measurement compression protocol, which is in general not true. The formal statement is as follows :
\begin{lemma}
{\bf Classical Message Compression with Quantum Side Information \\}\label{lem:SlepianWolf}
Given a classical quantum state of the form
\[
\sigma^{XB}\coloneqq\sum_{x}P_X(x)\ketbra{x}^X\otimes \sigma^B
\]
where Alice possesses the system $X$ and Bob possess the system $B$, there exists a CDC-QSI protocol with probability of error at most $\epsilon$
and the rate of communication $R_A$ from Alice to Bob is at least
\[
R_A \geq \phmax^{\eps}(X)-I_H^{\eps}(X:B)+O\left(\log\frac{1}{\eps}\right)
\]
where all entropic quantities are computed in terms of the state $\sigma^{XB}$.
\end{lemma}

\subsection{Centralised Measurement Compression}
In this section we briefly explain our strategy to design a protocol for centralised multi link measurement compression. Due to the composability of our CDC-QSI lemma with any measurement compression scheme, we combine these two protocols to achieve the bounds claimed in \cref{thm:CentralisedMeasurementCompression}. The technique that we use to design the compressed POVMs $\theta_k$ first requires the creation of a random codebook $\mathcal{C}\coloneqq \brak{X(k,\ell)}$ sampled iid from the distribution $P_X(x)\coloneqq \Tr[\rho\Lambda_x]$. We further require that, on expectation over choices of codebook, the following \emph{sample average state} is close to 
{\small
\[
\frac{1}{L}\sum_{\ell\in [L]}\frac{1}{P_X(x(k,l))}\sqrt{\rho}\Lambda_{x(k,\ell)}\sqrt{\rho}\overset{\eps}{\approx} \rho
\]
}
In the centralised case however, since we are forced to construct the two codebooks for the $X$ and $Y$ links independently from the distributions $P_X$ and $P_Y$, the \emph{sample average matrices} that we must consider are of the form
{\small
\[
\frac{1}{L_1\cdot L_2} \sum\limits_{\ell_1,\ell_2}\frac{1}{P_X(x(k_1,\ell_1))P_Y(y(k_2,\ell_2))}\sqrt{\rho}\lambda_{\substack{x(k_1,\ell_1) \\ y(k_2,\ell_2)}}\sqrt{\rho}
\]
}
Even though the above matrix may not even be a quantum state, we still require that in expectation, it should be close to $\rho$. We show this using our \emph{measure transformed sequential quantum covering lemma } :
\begin{lemma}{\bf Measure Transformed Sequential Covering Lemma \\}\label{lem:coveringlemma}
Suppose we are given a joint distribution $P_{XY}$ on classical alphabets $\mathcal{X}\otimes \mathcal{Y}$, with marginals $P_X$ and $P_Y$. Suppose we are also given the following quantum state:
{\small
\[
\rho^{XYE}\coloneqq \sum P_{XY}(x,y) \ket{x}\bra{x}^X \otimes \ket{y}\bra{y}^Y\otimes \rho_{x,y}
\]
\small
}
Let $\brak{x(1),x(2),\ldots, x(K)}$ and $\brak{y(1),y(2),\ldots, y(L)}$ be iid samples from the distribution $P_X\otimes P_Y$. Then 

{ \small
\[
\E_{\substack{x(1),x(2),\ldots, x(K) \\ y(1),y(2),\ldots, y(L)}}
\left\lVert
\frac{1}{K\cdot L}\sum\limits_{k,l}
\frac{P_{XY}(x(k),y(l))}{P_X(x(k))\cdot P_Y(y(l))}\rho_{x(k),y(l)}
-\sigma^E 
\right\rVert_1 \leq \epsilon
\]
}
whenever 
{\small
\begin{align*}
\log K > I_{\max}^{\epsilon}(X : E) \textup{ and }
\log L > I_{\max}^{\epsilon}(Y:XE)
\end{align*}
}
where 
{\small
\[
\sigma^E \coloneqq \sum\limits_{x,y} P_{XY}(x,y) \rho^E_{x,y}
\]
}
\end{lemma}

\subsection{Organisation}
The paper is organised as follows: In \cref{sec:entropicQuantities} we define the one shot quantities that we will need for the proof, and state some useful facts. In \cref{sec:technicalTools} we present the main technical tools that we will use in the proof of our measurement compression theorem, including the measure transformed sequential covering lemma \cref{subsec:coveringLemma}, a one shot operator inequality that arises from this covering lemma \cref{subsec:opineq} and the one shot classical message compression protocol with quantum side information \cref{subsec:SlepianWolf}. Then in \cref{sec:MeasurementCompression} we prove a version of our centralised multi link measurement compression theorem in the case when there is no side information available to the receiver Bob. In \cref{subsec:MCSlepianWolf}, we show how one can compose a measurement compression theorem with the protocol for classical message compression with quantum side information. Finally, we prove our full centralised multi link measurement compression theorem with side information in \cref{subsec:finalproof}. In \cref{sec:iid} we show how our one-shot bounds can be extended to the desired rate region in the asymptotic iid setting.
\section{Preliminaries: Relevant Quantities}\label{sec:entropicQuantities}

\subsection{Notation and Some Basics}

\begin{definition}[{\bf Quantum State}] A quantum state $\rho$ on some register $A$ is a positive semi-definite matrix with trace $1$.
\end{definition}

\begin{remark}
    The notation $\sigma \geq 0$, for a matrix $\sigma$ is used to denote the fact that $\sigma$ is positive semi-definite. More generally, $\rho \geq \sigma$ implies that the matrix $\rho-\sigma\geq 0$. This partial order of the positive semi-definite matrices is referred to as the Loewner order.
 \end{remark}

 \begin{definition}[{\bf Fidelity and Generalised Fidelity}]
     Given two quantum states $\rho$ and $\sigma$, the fidelity between the two states is defined as:
     \[
     F(\rho, \sigma)\coloneqq \norm{\sqrt{\rho}\sqrt{\sigma}}_1.
     \]
     The fidelity can be extended in a meaningful way to matrices which are sub-states, i.e. matrices $\rho$ and $\sigma$ such that $0 \leq \rho, \sigma, \leq I $, in the following way:
     \[
     \overline{F}(\rho, \sigma)\coloneqq F(\rho, \sigma)+\sqrt{(1-\Tr[\rho])(1-\Tr[\sigma])}.
     \]
     $\overline{F}(\cdot, \cdot )$ is referred to as the generalised fidelity.
 \end{definition}

 \begin{remark}
 The generalised fidelity was defined in \cite{dualitysmoothminmax}.
 \end{remark}

 \begin{definition}{\bf Purified Distance}
Given two quantum states $\rho^A$ and $\sigma^A$, the purified distance $P(\rho,\sigma)$ between the two states is defined as 
\[
P(\rho,\sigma)\coloneqq \sqrt{1-F^2(\rho,\sigma)}
\]
where $F(\cdot ~|~ \cdot)$ is the fidelity.
\end{definition}

\begin{definition}
    Given a quantum state $\rho^A$, we define the $\eps$-ball $\mathcal{B}^{\eps}(\rho)$ as:
    \[
    \mathcal{B}^{\eps}(\rho)\coloneqq \brak{\tau~:~0\leq \tau, \Tr\left[\tau\right]\leq 1, P(\tau, \rho)\leq \eps},
    \]
    where $P(\cdot, \cdot)$ is the purified distance on the space of sub-normalised states (see \cite{dualitysmoothminmax} for details).
\end{definition}

 \begin{definition}{\bf $\cdot$ Operation}
    Given an operator $M^{A\to B}$ and the operator $N^A$, we define the $\cdot$ operation as follows:
    \[
    M\cdot N\coloneqq MNM^{\dagger}.
    \]
\end{definition}
\subsection{Smooth Max Entropy}
In this section we introduce the one-shot entropic quantities which we will be  using in the subsequent sections to describe our protocols.

\begin{definition}[\textbf{Smoothed Support Max Entropy}]\label{def:hmaxtilde}
Given a quantum state $\rho^A$, let us denote its eigenvalues by $\lambda_1,\ldots \lambda_{\abs{\textup{supp}(\rho)}}$ (in ascending order) corresponding to the eigenvectors $ v_1, \ldots, v_{\abs{\textup{supp}(\rho)}}$. Let $\lambda_1, \ldots \lambda_k$ denote the smallest eigenvalues such that $\sum_{i} \lambda_i \leq \eps$. We define the {\bf smoothed support max entropy} of $\rho^A$ as 
\[
\thmax^{\eps}(A)_{\rho}\coloneqq \log \left(\abs{\textup{supp}(\rho)}-k\right).
\]
\end{definition}

\begin{definition}[\textbf{Smoothed Norm Max Entropy}]\label{def:hmax}
Given the setup of in Definition \ref{def:hmaxtilde}, we define the {\bf smoothed norm max entropy} of the state $\rho^A$ as
\[
\phmax^{\eps}(A)_{\rho}\coloneqq \log \frac{1}{\lambda_{k+1}}
\]
\end{definition}

\begin{definition}{{\bf (Conditional Smooth Max Entropy)}}
Given a bipartite quantum state $\rho^{AB}$, we define the {\bf smooth max entropy } as
\[
H_{\max}^{\eps}(A|B)_{\rho} \coloneqq \min\limits_{\rho'\in \mathcal{B}^{\eps}(\rho)}\max\limits_{\substack{\sigma^B\geq 0\\ \Tr[\sigma]=1}} 2\log F\left(\rho'^{AB}, \textup{\one}^A\otimes \sigma^B \right)
\]

\begin{fact}[\cite{CNB23}]\label{fact:tildemaxHrelation}
For any quantum state $\rho^A$ it holds that
\[
H_{\max}^{\eps}(A)_{\rho}\leq \thmax^{\eps}(A)_{\rho} \leq  \phmax^{\eps}(A)_{\rho} \leq \log\frac{\abs{A}}{\eps}
\]
\end{fact}

\begin{remark}
In this paper we will only use the \emph{smoothed norm max entropy} $\phmax^{\epsilon}$.
\end{remark}

\end{definition}

\subsection{Hypothesis Testing Relative Entropy}
\begin{definition}{\bf Smooth Hypothesis Testing Relative Entropy}
Given quantum state $\rho^{A}$ and $\sigma^A$, the smooth hypothesis testing relative entropy between $\rho$ and $\sigma$ is defines as
\[
D_{H}^{\eps}(\rho^A~||~\sigma^A)\coloneqq -\log \textsc{OPT}
\]
where $\textsc{OPT}$ is the optimum attained for the following semi-definite program
\begin{align*}
    \max &\Tr[\Pi \sigma] \\
    &\Tr[\Pi \rho] \geq 1-\eps \\
    & 0 \leq \Pi \leq \I 
\end{align*}
\end{definition}

\begin{definition}{\bf Hypothesis Testing Mutual Information}
Given the quantum state $\rho^{AB}$ , the hypothesis testing mutual information $I_{H}^{\eps}(A:B)$ is defined as
\[
I_{H}^{\eps}(A:B)_{\rho}\coloneqq D_{H}^{\eps}(\rho^{AB}~||~\rho^A\otimes \rho^B)
\]
\end{definition}
\begin{fact}
Given a classical quantum state 
\[
\sum_xP_X(x)\ketbra{x}^X\otimes \rho_x^B
\]
the operator $\Pi_{\textsc{opt}}$ which is the optimiser for $I_{H}^{\eps}(X:B)_{\rho}$ is of the form
\[
\Pi_{\textsc{opt}}=\sum_x\ketbra{x}^X\otimes \Pi_x^B
\]
where for each $x$, 
\[
0\leq \Pi_x\leq \I
\]
\end{fact}

\subsection{Smooth Max Relative Entropy}
\begin{definition}{\bf  Max Relative Entropy}
Given a quantum states $\rho^{A}$ and $\sigma^A$, the max relative entropy $D_{\max}(\rho^{A}~||~\sigma^A)$ is defined as 
\[
D_{\max}(\rho^A~||\sigma^A)\coloneqq \inf \brak{\lambda~|~\rho^A\leq 2^{\lambda}\sigma^A}
\]
\end{definition}

\begin{definition}{\bf Max Information}
Given a bipartite quantum state $\rho^{AB}$, the max information $I_{\max}(A : B)$ is defined as
\[
I_{\max}(A : B)\coloneqq D_{\max}(\rho^{AB}~||~\rho^A\otimes \rho^B).
\]
\end{definition}
\begin{definition}{\bf Smooth Max Information}\label{def:smoothMaxInfo} Given a state $\rho^{AB}$ the smooth max information $I_{\max}^{\epsilon}(A:B)_{\rho}$ is given by 
\[
I_{\max}^{\eps}(A:B)_{\rho} \coloneqq \inf\limits_{\rho'\in\mathcal{B}^{\eps}(\rho)} I_{\max}(A : B)_{\rho'}.
\]
\end{definition}

We will need a slightly perturbed version of this quantity to state our theorems. We call this the tilde smooth max information.

\begin{definition}{\bf Tilde Smooth Max Information \cite{WildeWiretap}}\label{def:tildeSmoothMaxInfo}
Given a state $\rho^{AB}$ the tilde smooth max information between $A$ and $B$ is defined as 
\[
\widetilde{I}_{\max}^{\eps}(B:A)_{\rho}\coloneqq \inf\limits_{\rho'^{AB} : P(\rho^{AB}, \rho'^{AB})\leq \epsilon} D_{\max}(\rho'^{AB}~||~\rho^A\otimes \rho'^{B})
\]
\end{definition}
We will also need the following quantity:
\begin{definition}{\bf First Argument Smoothed Max Information}
    Given a bipartite state $\rho^{AB}$, the first argument smoothed max information $J^{\eps}_{\max}(A : B)_{\rho}$ is defined as:
    \[
    J^{\eps}_{\max}(A : B)_{\rho}\coloneqq \inf\limits_{\rho'\in \mathcal{B}^{\eps}(\rho)} D_{\max}(\rho'^{AB}~||~\rho^A\otimes \rho^B).
    \]
\end{definition}
\begin{fact}{\bf \cite{WildeWiretap}} \label{fact:imaxepsimax}
For any state $\rho^{AB}$, $\epsilon>0$ and $0<\gamma<\epsilon$, the following is true 
\[
\widetilde{I}_{\max}^{\eps}(B:A)_{\rho} \leq J_{\max}^{\eps-\gamma}(A:B)_{\rho}+\log\frac{3}{\gamma^2}
\]
\end{fact}
\begin{lemma}\label{lem:SmoothMaxInfoEquiv}
        For any bipartite state $\rho^{AB}$ and $\eps>0$ it holds that:
        \[
        J^{2\eps^{1/4}}(A:B)_{\rho}\leq I_{\max}^{\eps}(A:B)_{\rho}+\log \frac{1}{1-4\sqrt{\eps}}.
        \]
    \end{lemma}

    The proof of this lemma is included in the Appendix \ref{sec:appendixA}.

\subsection{Other Useful Tools}

A version of the smoothed convex split lemma was proven in \cite{WildeWiretap}. However, that lemma was stated in terms of the tilde smooth max information (see Definition \ref{def:tildeSmoothMaxInfo}). Below we state a version of this important lemma with the more standard smooth max mutual information (see Definition \ref{def:smoothMaxInfo}). Although we state it as a fact, we included the proof of this version in Appendix \ref{sec:appendixB}, since it is somewhat different from the original.

\begin{fact}{\bf Smoothed Convex Split Lemma}\label{fact:smoothedConvexSplitLemma}

Let $\rho^{AB}$ be any quantum state and let $\tau^{A_1\ldots A_K B}$ be the so called convex split state, defined as 
\[
\tau^{A_1\ldots A_K B} \coloneqq \frac{1}{K}\sum\limits_{k \in [K]} \rho^{A_kB} \Otimes\limits  \rho^{A_{[K]\setminus k}} 
\]

Then for all
\[
\log K \geq I_{\max}^{\eps^4}(A:B)_{\rho}+O\big(\log \frac{1}{\eps}\big)
\]
the following holds true
\[
\norm{\tau^{A_1\ldots A_KB}-\rho^B\Otimes \rho^{A_{[K]}}}_1 \leq 20\eps
\]
where $\eps\in (0,1)$.
\end{fact}

\subsection{Quantum Asymptotic Equipartition Property}\label{subsec:QAEP}
We will find the following facts useful while extending our results to the asymptotic iid setting. We collectively refer to them as the Quantum Asymptotic Equipartition Property (QAEP). The proofs of these facts can be found in \cite{TomamichelHayashi, TomamichelTan, KeLi, QAEP}.
\begin{fact}
Given a classical probability distribution $P_X$ and some integer $n\in \mathbb{N}$, the following holds
\[
\lim\limits_{\eps\to 0} \lim\limits_{n\to \infty}\frac{1}{n} \phmax^{\eps}(X)=H(X)
\]
\end{fact}
\begin{fact}{\bf Asymptotic Equipartition Property}
Given the quantum states $\rho^{A}$ and $\sigma^A$, and some integer $n\in \mathbb{N}$, the following hold:
\begin{align*}
   &\lim\limits_{\eps\to 0} \lim\limits_{n\to \infty}\frac{1}{n} D_H^{\eps}(\rho^{\otimes n}~||~\sigma^{\otimes n}) = D(\rho~||~\sigma) \\
   & \lim\limits_{\eps\to 0} \lim\limits_{n\to \infty}\frac{1}{n} D_{\max}^{\eps}(\rho^{\otimes n}~||~\sigma^{\otimes n}) = D(\rho~||~\sigma)
\end{align*}
where $D(\rho~||~\sigma)$ is the quantum von Neumann relative entropy.
\end{fact}

\begin{fact}{\bf AEP for the Smooth Max Information}
    Given a quantum state $\rho^{AB}$, it holds that:
    \[
    \lim\limits_{\eps\to 0}\lim\limits_{n\to \infty} \frac{1}{n}I^{\eps}_{\max}\left(A^n : B^n\right)_{{\rho}^{\otimes n}} = I\left(A : B\right)_{\rho}.
    \]
    This result is implied by the results in \cite{Ciganovic} and the AEP in \cite{ReverseShannonTheorem}.
\end{fact}

\section{Technical Tools}\label{sec:technicalTools}
In this section we will prove the two main technical tools we will need to prove our centralised multi link measurement compression theorem: 
\begin{enumerate}
    \item the measure transformed sequential covering lemma,
    \item a new operator inequality implied by the covering lemma, and 
    \item the protocol for classical message compression with quantum side information.
\end{enumerate}
\subsection{A Measure Transformed Sequential Covering Lemma}\label{subsec:coveringLemma}
In this section we will prove \cref{lem:coveringlemma}. To do this we will first need to prove the following sequential convex split lemma:
\begin{theorem}{\bf Successive Cancellation Convex Split}\label{thm:SuccCancConvSplit}
Suppose we are given the tripartite state $\rho^{ABR}$. Define the convex split state as follows 
\[
\tau^{A_1\ldots A_K B_1 \ldots B_L R} \coloneqq \frac{1}{K\cdot L}\sum\limits_{k, \ell \in [K], [L]} \rho^{A_kB_{\ell} R} \Otimes \rho^{A_{[K]\setminus k}} \Otimes\limits \rho^{B_{[L]\setminus\ell}} 
\]
Then whenever 
 \begin{align*}
        \log K &\geq I_{\max}^{\eps^4}(A:BR)_{\rho} + O\big(\log \frac{1}{\eps}\big) \\
    \log L &\geq I_{\max}^{\eps^4}(B:R)_{\rho} + O\big(\log \frac{1}{\eps}\big)
\end{align*}
the following holds
\[
\norm{\tau^{A_1\ldots A_K B_1 \ldots B_L R}-\rho^{R}\Otimes\limits_{i\in [K]}\rho^{A_i}\Otimes\limits_{j\in [L]} \rho^{B_j}}_1 \leq 40\eps.
\]

\end{theorem}

\begin{proof}
Consider the expression
\begin{align}\label{eq1}
    &\norm{\tau^{A_1\ldots A_K B_1\ldots B_L R}-\rho^{R}\Otimes\limits_{i\in [K]}\rho^{A_i}\Otimes\limits_{j\in [L]} \rho^{B_j}}_1 \\
    \leq & \norm{\tau- \frac{1}{L}\sum\limits_{\ell \in [L]}\rho^{B_{\ell}R}\Otimes\rho^{B_{[L]\setminus \ell}}\Otimes\limits_{i\in [K]} \rho^{A_i} }_1+\norm{\frac{1}{L}\sum\limits_{\ell \in [L]}\rho^{B_{\ell}R}\Otimes\limits\rho^{B_{[L]\setminus \ell}}\Otimes\limits_{i\in [K]} \rho^{A_i}- \rho^{R}\Otimes\limits_{i\in [K]}\rho^{A_i}\Otimes\limits_{j\in [L]} \omega^{B_j}}_1
\end{align}
Let us first consider the first term in the RHS. Note that we can write this term as follows
\begin{align*}
    &\norm{\tau- \frac{1}{L}\sum\limits_{\ell \in [L]}\rho^{B_{\ell}R}\Otimes\limits_{j\neq \ell}\rho^{B_{[L]\setminus\ell}}\Otimes\limits_{i\in [K]} \rho^{A_i} }_1 \\
    \leq &\frac{1}{L}\sum\limits_{\ell \in [L]} \norm{\frac{1}{K}\sum\limits_{k\in [K]}\rho^{A_kB_{\ell}R}\Otimes\rho^{A_{[K]\setminus k}} \Otimes\rho^{B_{[L]\setminus \ell}}-\rho^{B_{\ell}R}\Otimes\rho^{B_{[L]\setminus\ell}}\Otimes\limits_{i\in [K]} \rho^{A_i}}_1 \\
    \intertext{Taking the systems $B_1\ldots B_{\ell-1}B_{\ell+1}\ldots B_{L}$ systems from each of the corresponding norm expressions outside the norm, we see that RHS}
    = & \frac{1}{L}\sum\limits_{\ell \in [L]} \norm{\frac{1}{K}\sum\limits_{k\in [K]}\rho^{A_kB_{\ell}R}\Otimes \rho^{A_{[K]\setminus k}} -\rho^{B_{\ell}R}\Otimes\limits_{i\in [K]} \rho^{A_i}}_1
\end{align*}
By the smoothed convex split lemma (\ref{fact:smoothedConvexSplitLemma}), for 
\[
\log K \geq I_{\max}^{\eps^4}(A:BR)_{\rho}+O\big(\log \frac{1}{\eps}\big)
\]
we see that, for all $\ell\in [L]$, the terms 
\[
\norm{\frac{1}{K}\sum\limits_{k\in [K]}\rho^{A_kB_{\ell}R}\Otimes \rho^{A_{[K]\setminus k}} -\rho^{B_{\ell}R}\Otimes\limits_{i\in [K]} \rho^{A_i}}_1 \leq 20\eps.
\]
We will now consider the second term in the RHS of \cref{eq1}. Notice that,
\begin{align*}
     & \norm{\frac{1}{L}\sum\limits_{\ell \in [L]}\rho^{B_{\ell}R}\Otimes\limits_{j\neq \ell}\rho^{B_{j}}\Otimes\limits_{i\in [K]} \rho^{A_k}- \rho^{R}\Otimes\limits_{i\in [K]}\rho^{A_i}\Otimes\limits_{j\in [L]} \rho^{B_j}}_1 \\
     = & \norm{\frac{1}{L}\sum\limits_{\ell \in [L]}\rho^{B_{\ell}R}\Otimes\limits_{j\neq \ell}\rho^{B_{j}}- \rho^{R}\Otimes\limits_{j\in [L]} \rho^{B_j}}_1
\end{align*}
Then by another application of the convex split lemma, for all 
\[
\log L \geq I_{\max}^{\eps^4}(B : R)_{\rho} + O\big(\log \frac{1}{\eps}\big)
\]
the following holds
\[
\norm{\frac{1}{L}\sum\limits_{\ell \in [L]}\rho^{B_{\ell}R}\Otimes\limits_{j\neq \ell}\rho^{B_{j}}- \rho^{R}\Otimes\limits_{j\in [L]} \rho^{B_j}}_1 \leq 20\eps.
\]
Collating all these arguments we see that 
\begin{align*}
    \norm{\tau^{A_1\ldots A_K B_1\ldots B_L R}-\rho^{R}\Otimes\limits_{i\in [K]}\rho^{A_i}\Otimes\limits_{j\in [L]} \rho^{B_j}}_1 \leq 40\eps.
\end{align*}
This concludes the proof.
\end{proof}
We will now use \cref{thm:SuccCancConvSplit} to prove \cref{lem:coveringlemma} as a corollary. We restate the lemma below for convenience.
\begin{corollary}{\bf Measure Transformed Sequential Covering Lemma \\}\label{corol:MeasTransCoveringLemma}
Suppose we are given a joint distribution $P_{XY}$ on classical alphabets $\mathcal{X}\otimes \mathcal{Y}$, with marginals $P_X$ and $P_Y$. Suppose we are also given the following quantum state:
\[
\rho^{XYE}\coloneqq \sum P_{XY}(x,y) \ket{x}\bra{x}^X \otimes \ket{y}\bra{y}^Y\otimes \rho_{x,y}
\]
Let $\brak{x(1),x(2),\ldots, x(K)}$ and $\brak{y(1),y(2),\ldots, y(L)}$ be iid samples from the distribution $P_X\otimes P_Y$. Then 
\[
\E_{\substack{x(1),x(2),\ldots, x(K) \\ y(1),y(2),\ldots, y(L)}}\norm{\frac{1}{K\cdot L}\sum\limits_{k,l}\frac{P_{XY}(x(k),y(l))}{P_X(x(k)\cdot P_Y(y(l)))}\rho_{x(k),y(l)}-\sigma^E}_1 \leq O(\epsilon)
\]
whenever 
\begin{align*}
\log K &> I_{\max}^{\epsilon^4}(X : E) + O\big(\log \frac{1}{\eps}\big)\\
\log L &> I_{\max}^{\epsilon^4}(Y:XE) + O\big(\log \frac{1}{\eps}\big)
\end{align*}
where 
\[
\sigma^E \coloneqq \sum\limits_{x,y} P_{XY}(x,y) \rho^E_{x,y}
\]
\end{corollary}

\begin{proof}
We apply \cref{thm:SuccCancConvSplit} after instantiating terms appropriately. For our case, the classical systems $X$ and $Y$ play the roles of the systems $A$ and $B$ in \cref{thm:SuccCancConvSplit}. Then, recall that
\[
\tau^{X_1\ldots X_K Y_1 \ldots Y_L R} \coloneqq \frac{1}{K\cdot L}\sum\limits_{k, \ell \in [K], [L]} \rho^{X_kY_{\ell} R} \Otimes \rho^{X_{[K]\setminus k}} \Otimes\limits \rho^{Y_{[L]\setminus\ell}} 
\]
where 
\begin{align*}
    \rho^X \coloneqq &\sum\limits_{x}P_X(x)\ketbra{x}^X \\
    \rho^Y \coloneqq &\sum\limits_{y}P_Y(y)\ketbra{y}^Y
\end{align*}
Note that the states $\rho^X$ and $\rho^Y$ are marginals of the control state $\rho^{XYE}$.
Let us consider one term in the expansion of $\tau$, for some fixed $k\in [K]$ and $\ell \in [L]$
\[
\rho^{X_kY_{\ell}R}\bigotimes \rho^{X_{[K]\setminus k}}\bigotimes \rho^{Y_{[L]\setminus \ell}} \tag{1}\label{eq:fixed}
\]
To describe this fixed term we use the following convention : we write $x_i$ to denote a sample from the system $X_i$, and similarly for $y_j$. Let us fix a certain setting of these samples 
\[
\prod\limits_{i\in [K]} \ketbra{x_i}^{X_i} \otimes \prod \limits_{j\in [L]}\ketbra{y_j}^{Y_j}
\]
Since the $X_k$ and the $Y_{\ell}$ systems are entangled with the $R$ system for the state in \cref{eq:fixed}, the term that appears in the expansion of the state in \cref{eq:fixed} is
\[
P_{XY}(x_k,y_{\ell})\ketbra{x_k}^{X_k}\otimes \ketbra{y_{\ell}}^{Y_{\ell}}\otimes \prod\limits_{\substack{i\in [K],~ j\in [L] \\ i\neq k,~ j\neq \ell}}P_X(x_i)P_Y(y_j) \otimes \ketbra{x_i}^{X_i} \ketbra{y_j}^{Y_j}\otimes \rho^{R}_{x_k,y_{\ell}} \tag{2}\label{eq:2}
\]
The full expansion of the state in \cref{eq:fixed} is a sum over all strings $x_1,x_2,\ldots, x_K$ and $y_1,y_2,\ldots, y_L$, of terms like the one in \cref{eq:2}. Now, when we sum over all fixed states as in \cref{eq:fixed} to get $\tau$, we can group together those matrices of the kind in \cref{eq:2} which have the same string $x_1,x_2,\ldots, x_K, y_1,y_2,\ldots, y_L$. Note that this choice of fixed string fixes the pure states on the systems $X_1X_2\ldots X_K Y_1Y_2\ldots Y_L$. However, the corresponding state on the system $R$ will be of the following form
\[
\frac{1}{K\cdot L}\sum\limits_{k,\ell} P_{XY}(x_k,y_{\ell})~\prod\limits_{\substack{i\neq k \\ j\neq \ell}}P_X(x_i)\cdot P_Y(y_j) \rho^{R}_{x_k,y_{\ell}}
\]
the above argument, along with \cref{thm:SuccCancConvSplit}, essentially implies the following :
\begin{align*}
    &\norm{\tau^{X_1X_2\ldots X_KY_1Y_2\ldots Y_LR}-\sigma^R\bigotimes\limits_{i\in [K]}\rho^{X_i}\bigotimes\limits_{j\in [L]}\rho^{Y_j}}_1 \\
    =&\left\lVert\sum\limits_{\substack{x_1, x_2,\ldots x_K \\ y_1, y_2, \ldots y_L}}x_1, x_2,\ldots x_k\otimes y_1, y_2, \ldots y_L\otimes \frac{1}{K\cdot L}\sum\limits_{k,\ell} P_{XY}(x_k,y_{\ell})~\prod\limits_{\substack{i\neq k \\ j\neq \ell}}P_X(x_i)\cdot P_Y(y_j) \rho^{R}_{x_k,y_{\ell}}\right.  \\
    -&\left.\sum\limits_{\substack{x_1,x_2,\ldots x_K \\ y_1, y_2, \ldots y_L}}x_1, x_2,\ldots x_k\otimes y_1, y_2, \ldots y_L\otimes \prod_{\substack{i\in [K],~j\in [L]}}P_X(x_i)\cdot P_Y(y_j) \sigma^R \right\rVert_1 \\
    \intertext{where we have omitted the ket bra notation on the classical string for brevity. Then, using the block diagonal nature of these matrices, we see that the above expression is}
    =& \sum\limits_{\substack{x_1, x_2,\ldots x_K \\ y_1, y_2, \ldots y_L}}\prod_{\substack{i\in [K],~j\in [L]}}P_X(x_i)\cdot P_Y(y_j) \norm{\frac{1}{K\cdot L}\sum_{k,\ell}\frac{P_{XY}(x_k,y_{\ell})}{P_X(x_k)\cdot P_Y(y_{\ell})}\rho^R_{x_k,y_{\ell}}-\sigma^R}_1 \\
    \intertext{We now change notation from $x_i$ and $y_j$ to $x(i)$ and $y(j)$ to emphasise the fact that they are the $i$-th and $j$-th samples drawn from the distributions $P_X$ and $P_Y$ respectively. Then, invoking \cref{thm:SuccCancConvSplit}, we see that the above expression is }
    =&\E\limits_{{\substack{x(1), x(2),\ldots x(K) \\ y(1), y(2), \ldots y(L)}}} \norm{\frac{1}{K\cdot L}\sum_{k,\ell}\frac{P_{XY}(x(k),y({\ell}))}{P_X(x(k))\cdot P_Y(y({\ell}))}\rho^R_{x(k),y(\ell)}-\sigma^R}_1 \\
    \leq &~ O(\eps) \\
    \intertext{for appropriately chosen values of $K$ and $L$.}
\end{align*}
This concludes the proof of the claim.
\end{proof}

\subsection{An Operator Inequality}\label{subsec:opineq}

In this section we prove an operator inequality that we will find useful in proving our measurement compression theorem.

\begin{lemma}{\bf Operator Inequality for the Covering Lemma}\label{lem:opineq}
Suppose we are given some states $\rho_1^A, \rho_2^A, \ldots, \rho_L^A$ and a distribution $P(i)$ on the set of indices $[L]$ such that
\[
\norm{\sum\limits_{i}P(i)\rho_i^A-\rho^A}_1 \leq \epsilon
\]
for a given average state $\rho$. Then there exist a good subset $\good\subset [L]$ such that 
\[
\Pr_{P}[\good] \geq 1-O(\epsilon^{1/4})
\]
and quantum states $\brak{\rho'_i~|~i\in \good}$ such that
\begin{align*}
    \norm{\rho^{'A}_i-\rho_i}_1 \leq 2\epsilon^{1/4} \\
    \intertext{and} 
    \sum_{i\in \good}P(i)\rho^{'A}_i \leq (1+\epsilon^{1/4}) \rho^A
\end{align*}
\end{lemma}
\begin{proof}
Let the state $\ket{\rho}^{AB}$ be a purification of $\rho^A$. Similarly, let $\ket{\rho_i}^{AB}$ denote purifications of the states $\rho_i^A$. Consider the purification :
\[
\ket{\psi}^{ABC}\coloneqq\sum_i\sqrt{P(i)}\ket{\rho_i}^{AB}\ket{i}^C
\]
of the sample average state
\[
\sum_i P(i)\rho_i^A
\]
Then by Uhlmann's theorem and the closeness in $1$-norm of the sample average state to $\rho^A$, we see that there exists a pure state $\ket{\varphi}^{ABC}$ such that
\[
\norm{\varphi-\psi}_1 \leq 2\sqrt{\epsilon}
\]
We will expand $\ket{\varphi}^{ABC}$ by fixing the computational basis on $C$. This expansion can be written in the following form :
\[
\ket{\varphi}^{ABC}= \sum_i\ket{v_i}^{AB}\ket{i}^C
\]
where $\ket{v_i}$'s are some vectors on which we make no assumptions. We claim that each vector $\ket{v_i}^{AB}$ has length at most $1$. This is not hard to see, since taking the inner product of $\ket{\varphi}^{ABC}$ with itself, we see that
\[
\sum_i \braket{v_i|v_i} = 1
\]
The above observation can be used to define a distribution $Q(i)$ on $[L]$ in the following way :
\[
Q(i)\coloneqq \braket{v_i|v_i}
\]
We also define vectors $\ket{\rho'_i}^{AB}$ by normalising the $\ket{v_i}$'s :
\[
\ket{\rho'_i}^{AB}\coloneqq \frac{1}{\sqrt{Q(i)}}\ket{v_i}^{AB}
\]
This allows us to express $\ket{\varphi}^{ABC}$ as follows :
\[
\ket{\varphi}^{ABC}=\sum_i \sqrt{Q(i)}\ket{\rho'_i}^{AB}\ket{i}^{C}
\]
Note that by definition each $\ket{\rho'_i}^{AB}$ is a vector of length $1$, and hence by the Schmidt decomposition and tracing out the system $B$, can be seen as a purification of some quantum state $\rho^{'A}_i$. \\
\begin{claim}\label{claim:index}
There exists a subset of $[L]$ of probability (under $P$) of at least $1-O(\eps^{1/4})$ on which 
\[
P(i)\leq (1+\eps^{1/4}) Q(i)
\]
\end{claim}
\begin{proof}
Note that by the $2\sqrt{\eps}$ closeness of $\varphi$ and $\psi$ and the monotonicity of $1$-norm :
\[
\sum_iQ(i)\left\lvert 1-\frac{P(i)}{Q(i)}\right\rvert \leq 2\sqrt{\eps}
\]
Define the set
\[
\goodone\coloneqq \brak{i~|~\left\lvert 1- \frac{P(i)}{Q(i)}\right\rvert\leq \epsilon^{1/4}}
\]
From Markov's inequality, we see that 
\[
\Pr\limits_{Q}\left[\goodone^c\right] \leq 2\eps^{1/4}
\]
Since $\norm{Q-P}_1 \leq 2\sqrt{\eps}$, this implies that
\[
\Pr\limits_{P}\left[\goodone^c\right]\leq O(\eps^{1/4})
\]
Therefore for all $i\in \goodone$, we have that
\[
P(i)\leq (1+\eps^{1/4})Q(i)
\]
This proves the claim.
\end{proof}

\begin{claim}
There exists a subset of $[L]$ of probability under $P$ at least $1-O(\eps^{1/4})$ such that 
\[
\norm{\rho'_i-\rho_i}_1\leq 2\eps^{1/4}
\]
\end{claim}
\begin{proof}
Consider the expression
\[
\norm{\varphi-\psi}_1 \leq \sqrt{\eps}
\]
We will measure the matrices inside the $1$-norm along the computational basis on the system $C$. This produces states that are block diagonal. Appealing to the monotonicity of the $1$-norm we see that this implies
\[
\sum_i Q(i)\norm{\rho'_i-\frac{P(i)}{Q(i)}\rho_i}_1 \leq 2\sqrt{\eps}
\]
Define the set 
\[
\close \coloneqq \left\{i~|~\norm{\rho'_i-\frac{P(i)}{Q(i)}\rho_i}_1 \leq \eps^{1/4}\right\}
\]
By arguments similar to those used in \cref{claim:index}, we see that
\[
\Pr\limits_{P}\left[\close^c\right]\leq O(\eps^{1/4})
\]
Then, recalling the definition of the set \goodone ~from \cref{claim:index}, we see that
\[
\Pr\limits_{P}\left[\goodone\bigcap \close\right] \geq 1-O(\eps^{1/4})
\]
Define
\[
\good \coloneqq \goodone \bigcap \close
\]
Then, for all $i\in \good$ we see that
\[
\begin{aligned}
\norm{\rho'_i-\rho_i}_1 &\leq \norm{\rho'_i-\frac{P(i)}{Q(i)}\rho_i}+\norm{\frac{P(i)}{Q(i)}\rho_i-\rho_i}_1 \\
&\leq \eps^{1/4}+\eps^{1/4} \\
&=2\eps^{1/4}
\end{aligned}
\]
\end{proof}

We now define $\ket{\Tilde{\varphi}}$ by throwing away those indices from the expansion of $\ket{\varphi}$ which are not in $\good$.
\[
\ket{\Tilde{\varphi}}^{ABC}\coloneqq \sum\limits_{i\in \good}\sqrt{Q(i)}\ket{\rho'_i}^{AB}\ket{i}
\]
Then note that
\[
\Tr_{BC}\left[\Tilde{\varphi}\right] = \sum\limits_{i\in \good}Q(i)\rho^{'A}_i \leq \sum\limits_{i\in [L]}Q(i)\rho^{'A}_i =\rho
\]
Finally, by using the properties of \good, we observe that
\begin{align*}
    \sum\limits_{i\in \good}P(i)\rho^{'A}_i &\leq (1+\eps^{1/4})\sum\limits_{i\in \good}Q(i)\rho^{'A}_i \\
    &\leq (1+\eps^{1/4})\rho
\end{align*}
This concludes the proof.
\end{proof}

\begin{corollary}\label{corol:MeasureTransOpIneq}
Suppose we are given some positive semi-definite matrices $\sigma_1^A, \sigma_2^A, \ldots, \sigma_L^A$ and a distribution $P(i)$ on the set of indices $[L]$ such that
\[
\norm{\sum\limits_{i}P(i)\sigma_i^A-\rho^A}_1 \leq \epsilon
\]
for a given average state $\rho$. Define
\[
P'(i)\coloneqq \frac{P(i)\cdot \Tr[\sigma_i]}{\sum\limits_{i}P(i)\cdot \Tr[\sigma_i]}
\]
Then there exist a good subset $\good\subset [L]$ such that 
\[
\Pr_{P'}[\good] \geq 1-O(\epsilon^{1/4})
\]
and quantum states $\brak{\rho'_i~|~i\in \good}$ such that
\begin{align*}
    \norm{\rho^{'A}_i-\frac{\sigma_i}{\Tr[\sigma_i]}}_1 \leq 2\epsilon^{1/4} \\
    \intertext{and} 
    \sum_{i\in \good}P(i)\cdot \Tr[\sigma_i]\cdot\rho^{'A}_i \leq (1+O(\epsilon^{1/4})) \rho^A
\end{align*}
\end{corollary}

\begin{proof}
Firstly note that from the hypothesis given in the statement, using monotonicity of the $1$-norm
\[
\norm{\sum\limits_{i}P(i)\cdot \Tr[\sigma_i]-1}_1\leq \eps
\]
which implies that 
\[
1-\eps \leq \sum\limits_{i}P(i)\cdot \Tr[\sigma_i] \leq 1+\eps
\]
Let us define the distribution 
\[
P'(i)\coloneqq \frac{P(i)\cdot \Tr[\sigma_i]}{\sum\limits_{i}P(i)\cdot \Tr[\sigma_i]}
\]
We also define 
\[
\rho_i\coloneqq \frac{\sigma_i}{\Tr[\sigma_i]}
\]
We will need the following fact:

\begin{fact}
Given a state $\rho$ and a positive semi-definite matrix $\sigma$, the following holds:
\[
\norm{\rho-\frac{\sigma}{\Tr[\sigma]}}_1 \leq 2\norm{\rho-\sigma}_1
\]
\end{fact}
Using this fact we can now rewrite the condition in the theorem statement as:
\[
\norm{\sum\limits_{i}P'(i)\rho_i-\rho}_1 \leq 2\eps
\]
Then, invoking \cref{lem:opineq}, we can infer the existence of a set of indices \good such that
\[
\Pr\limits_{P'}[\good]\geq 1-O(\eps^{1/4})
\]
and states $\rho'_i$ such that
\[
\norm{\rho'_i-\rho_i}_1\leq O(\eps^{1/4})
\]
and
\[
\sum\limits_{i\in \good} P'(i)\rho'_i \leq (1+O(\eps^{1/4}))\rho
\]
Recall that 
\[
P'(i)=\frac{P(i)\cdot \Tr[\sigma_i]}{\sum\limits_{i}P(i)\cdot \Tr[\sigma_i]} \geq \frac{1}{1+\eps} P(i)\cdot \Tr[\sigma_i]
\]
Therefore, we conclude that
\[
    \sum_{i\in \good}P(i)\cdot \Tr[\sigma_i]\cdot\rho^{'A}_i \leq (1+O(\epsilon^{1/4})) \rho^A
\]
\end{proof}

\subsection{Classical Message Compression with Quantum Side Information}\label{subsec:SlepianWolf}
In this section we will prove \cref{lem:SlepianWolf}. To define the above task, abbreviated as CQSI, we consider two parties Alice and Bob and a classical quantum control state of the form
\[
\rho^{XB}\coloneqq \sum_x P_X(x)\ketbra{x}^X\otimes \rho_x^B
\]
where the classical system $X$ belongs to Alice and the quantum system $B$ belongs to Bob. Alice and Bob also share a forward noiseless classical channel from Alice to Bob. Alice wishes to send the contents of the classical register to Bob using as little classical communication as possible.

To be precise, Alice and Bob wish to create the state $\sigma^{X\hat{X}B}$ via classical communication, where the system $X$ belongs to Alice and $\hat{X}B$ belongs to Bob such that
\[
\norm{\rho^{X\hat{X}B}-\sigma^{X\hat{X}B}}_1\leq \eps
\]
where we define
\[
\rho^{X\hat{X}B}\coloneqq \sum_xP_X(x)\ketbra{x}^X\otimes \ketbra{x}^{\hat{X}}\otimes \rho_x^B
\]
We also allow Alice and Bob to share public random coins and they are allowed to use private coin algorithms for encoding and decoding. We wish to minimise the amount of classical communication from Alice to Bob. For convenience we restate \cref{lem:SlepianWolf} below.
\begin{theorem} \label{thm:SlepianWolf}
Given the control state
\[
\rho^{XB}= \sum_xP_X(x)\ketbra{x}^X\otimes \rho_x^B
\]
there exists a classical message compression protocol with quantum side information whenever the rate of communication $R$ satisfies
\[
R \geq \phmax^{\eps}(X)-I_H^{\eps}(X:B)_{\rho}+O\big(\log \frac{1}{\eps}\big)
\]
\end{theorem}
\begin{proof}
The rough idea is as follows :
\begin{enumerate}
    \item Alice hashes the set $\mathcal{X}$ into a smaller set $\mathcal{L}$ using a $2$-universal hash family $\mathcal{F}$,  where each $f\in \mathcal{F}$ maps $\mathcal{X}\to \mathcal{L}$.
    \item Each index $\ell\in \mathcal{L}$ then corresponds to a subset of symbols, usually called a 'bucket', which is essentially the pre-image of $\ell$ with respect to some randomly chosen hash function $f\in \mathcal{F}$.
    \item Alice and Bob are both provided the description of this randomly chosen has function before the protocol starts.
    \item One can imagine that Alice never receives a symbol which has low probability. Thus, we can essentially discard those $x$'s of low probability, which have at most $\eps$-mass under the distribution $P_X$. This implies that we can imagine that Alice receives her symbols from the sub-distribution $P'(x)$ instead of $P(x)$.
    \item Upon receiving the symbol $x$, Alice sends the hash $f(x)$ to Bob via the classical noiseless channel.
    \item Bob then knows the subset $f^{-1}(x)\subset \mathcal{X}$. To decode the correct $x$ sent by Alice, Bob does a measurement on his quantum system $B$. 
    \item To be able to distinguish among the members $x'\in f^{-1}(x)$, the quantum states $\rho_{x'}$ need to be sufficiently far apart.
    \item We show that this condition holds as long as 
    \[
    \abs{f^{-1}(x)}\leq 2^{I_{H}^{\eps}(X:B)_{\rho}}
    \]
    \item Recall that the support of $P'$ is of size at most $2^{\phmax^{\eps}(X)}$. Thus, the hash function essentially divides this set into $2^{\phmax^{\eps}(X)-I_H^{\eps}(X:B)_{\rho}}$ buckets. This concludes the protocol.
\end{enumerate}
The reason why the above protocol works is the 'random codebooks' created by the hash function. Without this randomness, we could  partition the support of $P'$ arbitrarily. But then we would not be able to guarantee that the symbols in the pre-image of any hash is distinguishable on average over the choice of the hash function. \\
\vspace{2mm}  \\ {\large {\bf Bob's Decoding} \\}
We wish to analyse the probability of a decoding error. However, we will not work with the distribution $P_X$ but with the sub-distribution $P'_X$. To see that this only incurs an extra $\eps$ error, note that
\begin{align*}
    \Pr[\textup{decoding error}] &= \sum_x P_X(x)\cdot \Pr[\textup{decoding error}~|~x] \\
    & = \sum_{x\in \textup{supp}(P')}P_X(x)\cdot \Pr[\textup{decoding error}~|~x] + \sum_{x\notin \textup{supp}(P')}P_X(X)(x)\cdot \Pr[\textup{decoding error}~|~x] \\
    & \leq \sum_{x}P'_X(x)\cdot \Pr[\textup{decoding error}~|~x] + \eps
\end{align*}
We first define the bucket corresponding to the the hash function $f$ and the hash $\ell$ as 
\[
\mathcal{A}(f,\ell)\coloneqq \brak{x~|~f(x)=\ell, x\in \textup{supp}(P')}
\]
Let us denote the elements in $\mathcal{A}(f,\ell)$ as $\brak{a_1^{\ell},a_2^{\ell},\ldots, a_{\abs{\mathcal{A}}}^{\ell}}$.
Now consider the operator $\Pi_{\textsc{opt}}$ from the definition of $I_{H}^{\eps}(X:B)_{\rho}$ and the associated operators $\Pi_x$. Upon receiving the has $\ell$, Bob sequentially measures his system $B$ with the operators $\Pi_{a_i^{\ell}}$.

We wish to analyse the probability of a decoding error. However, we will not work with the distribution $P_X$ but with the sub-distribution $P'_X$. To see that this only incurs an extra $\eps$ error, note that
\begin{align*}
    \Pr[\textup{decoding error}] &= \sum_x P_X(x)\cdot \Pr[\textup{decoding error}~|~x] \\
    & = \sum_{x\in \textup{supp}(P')}P_X(x)\cdot \Pr[\textup{decoding error}~|~x] + \sum_{x\notin \textup{supp}(P')}P_X(X)(x)\cdot \Pr[\textup{decoding error}~|~x] \\
    & \leq \sum_{x}P'_X(x)\cdot \Pr[\textup{decoding error}~|~x] + \eps
\end{align*}
Suppose that when the symbol $x$ is sent, the corresponding index of this symbol in the set $\mathcal{A}(f,\ell)$ is $a_m^{\ell}$. Then, conditioned on Alice having received $x$, the probability of incorrect decoding is given by
\begin{align*}
&1-\Tr\left[\Pi_{a_m^{\ell}}(\I-\Pi_{a_{m-1}^{\ell}})\ldots (\I-\Pi_{a_1^{\ell}})\cdot \rho_{a_m^{\ell}}\right]\\
=& \Tr\left[\rho_{a_m^{\ell}}\right]-\Tr\left[\Pi_{a_m^{\ell}}(\I-\Pi_{a_{m-1}^{\ell}})\ldots (\I-\Pi_{a_1^{\ell}})\cdot \rho_{a_m^{\ell}}\right] \\
\intertext{Using Sen's non-commutative union bound, the above expression can by bounded by}
\leq &\sqrt{\Tr\left[(\I-\Pi_{a_m^{\ell}}) \rho_{a_{m}^{\ell}}\right]+\sum\limits_{i=1}^{m-1}\Tr\left[\Pi_{a_i^{\ell}} \rho_{a_{m}^{\ell}}\right]}
\end{align*}
Now notice that, the sets $\mathcal{A}(f,\ell)$ form a disjoint cover of the set $\textsc{supp}\left(P'_X\right)$ over the indices $\ell$. Thus, taking an average over the elements of the set $\bigcup\limits_{\ell}\mathcal{A}(f,\ell)$ is the same as taking an average over the set $\textsc{supp}\left(P'_X\right)$. Using this observation along with the concavity of the square root, we see that the average error probability over choices of $x$ is at most
\begin{align*}
    &\sqrt{\sum_{\ell\in \mathcal{L}}\sum_{a_m^{\ell}\in \mathcal{A}(f,l)}P'_X(a_m^{\ell})\left(\Tr\left[(\I-\Pi_{a_m^{\ell}}) \rho_{a_{m}^{\ell}}\right]+\sum\limits_{i=1}^{m-1}\Tr\left[\Pi_{a_i^{\ell}} \rho_{a_{m}^{\ell}}\right]\right)} \\
    = &\sqrt{\sum_xP'_X(x)\Tr\left[(\I-\Pi_{x}) \rho_{x}\right]+\sum_{\ell\in \mathcal{L}}\sum_{a_m^{\ell}\in \mathcal{A}(f,l)}P'_X(a_m^{\ell})\sum\limits_{i=1}^{m-1}\Tr\left[\Pi_{a_i^{\ell}} \rho_{a_{m}^{\ell}}\right]}
\end{align*}
The first term inside the square root is at most $\eps$, by the property of $\Pi_{\textsc{opt}}$ that

\begin{align*}
\Tr\left[ \Pi_{\textsc{opt}}\sum_xP_X(x)\ketbra{x}^X\otimes\rho_x\right] &\geq 1-\eps \\
\intertext{which implies that}
\Tr\left[\sum_x P_X(x)\ketbra{x}^X\otimes \Pi_x\rho_x\right]&\geq 1-\eps
\end{align*}
It is easy to see from the above that
\[
\begin{aligned}
&\sum_xP'_X(x)\Tr\left[(\I-\Pi_{x}) \rho_{x}\right] \\
\leq &\sum_xP_X(x)\Tr\left[(\I-\Pi_{x}) \rho_{x}\right] \\
\leq & \eps
\end{aligned}
\]
To analyse the second term inside the square root, consider the following :
\begin{align*}
    &\sum_{\ell\in \mathcal{L}}\sum_{a_m^{\ell}\in \mathcal{A}(f,l)}P'_X(a_m^{\ell})\sum\limits_{i=1}^{m-1}\Tr\left[\Pi_{a_i^{\ell}} \rho_{a_{m}^{\ell}}\right] \\
    \leq &\sum_{\ell\in \mathcal{L}}\sum_{a_m^{\ell}\in \mathcal{A}(f,l)}P'_X(a_m^{\ell})\sum\limits_{i\neq m}\Tr\left[\Pi_{a_i^{\ell}} \rho_{a_{m}^{\ell}}\right] \\
    =& \sum_x P'_X(x) \sum_{\substack{x'\neq x \\ x'\in \textsc{supp}(P')}}I_{\brak{f(x')= f(x)}} \Tr\left[\Pi_{x'}\rho_x\right]
\end{align*}
where $I_{x'\neq x}$is the indicator for when $f(x')= f(x)$. We will now take an expectation over the choice of the hash function $f$. Note that the above term is inside a square root, so to do this we use the concavity if square root :
\begin{align*}
   &\E_f\left[ \sum_x P'_X(x) \sum_{x'\neq x}I_{\brak{x'\neq x}} \Tr\left[\Pi_{x'}\rho_x\right]\right] \\
   = & \sum_x P'_X(x) \sum_{\substack{x'\neq x \\ x'\in \textsc{supp}(P')}}\E_f\left[I_{\brak{f(x')= f(x)}}\right] \Tr\left[\Pi_{x'}\rho_x\right] \\
   =& \sum_x P'_X(x) \sum_{\substack{x'\neq x\\ x'\in \textsc{supp}(P')}}\Pr[f(x')= f(x)] \Tr\left[\Pi_{x'}\rho_x\right] \\
   \leq & 2^{-R}\sum_x P'_X(x) \sum_{\substack{x'\neq x\\ x'\in \textsc{supp}(P')}}\Tr\left[\Pi_{x'}\rho_x\right] \\
   \leq & 2^{-R}\sum_x P'_X(x) \sum_{x'\in \textsc{supp}(P')}\Tr\left[\Pi_{x'}\rho_x\right] \\
   \intertext{To bound this term, we multiply and divide by $P'_X(x')$ inside the second summation. This shows us that the above expression is equal to}
   =& 2^{-R}\sum_x P'_X(x) \sum_{x'\in \textsc{supp}(P')}\frac{P'_X(x')}{P'_X(x')}\Tr\left[\Pi_{x'}\rho_x\right] \\
   \leq & 2^{-R}\cdot 2^{\phmax^{\eps}(X)} \sum_xP'_X(x)\Tr\left[\sum_{x'\in \textsc{supp}(P')}\left(P'_X(x')\Pi_{x'}\right)\rho_x\right]
\end{align*}
where we have used Definition \ref{def:hmax} to upper bound each $\frac{1}{P'_X(x')}$ term by $2^{\phmax^{\eps}(X)}$. We will now switch back to the distribution $P_X$ by adding the terms corresponding to the $x$'s which not in the support of $P'_X$. This implies that the above expression can be upper bounded by 
\begin{align*}
    \leq & 2^{-R}\cdot 2^{\phmax^{\eps}(X)} \sum_x P_X(x)\Tr\left[\sum_{x'}\left(P_X(x')\Pi_{x'}\right)\rho_x\right] \\
    =& 2^{-R+\phmax^{\eps}(X)}\Tr\left[\left(\sum_{x'}P_X(x')\Pi_{x'}\right)\left(\sum_x P_X(x)\rho_x\right)\right] \\
    =& 2^{-R+\phmax^{\eps}(X)}\Tr\left[\left(\sum_{x'}P_X(x')\ketbra{x'}^X\otimes\Pi_{x'}^B\right)\I^X\otimes\left(\sum_x P_X(x) \rho_x^B\right)\right] \\
    =& 2^{-R+\phmax^{\eps}(X)}\Tr\left[\left(\sum_{x'}\ketbra{x'}^X\otimes\Pi_{x'}^B\right)\left(\sum_{x''}P_X(x'')\ketbra{x''}^X\right)\otimes\left(\sum_x P_X(x) \rho_x^B\right)\right] \\
    =& 2^{-R+\phmax^{\eps}(X)}\Tr\left[ \Pi_{\textsc{opt}}~\rho^X\otimes \rho^B\right] \\
    \leq & 2^{-R+\phmax^{\eps}(X)-I_{H}^{\eps}(X:B)_{\rho}}
\end{align*}
Thus, this shows that as long as 
\[
R \geq \phmax^{\eps}(X)-I_{H}^{\eps}(X:B)_{\rho}-\log \eps
\]
the average decoding error over choices of $x$ and the hash function $f$ is at most $\sqrt{2\eps}+\eps$.

To finish the proof, consider the left polar decomposition of the operator
\[
\Pi_{a_m^{\ell}}(\I-\Pi_{a_{m-1}^{\ell}})\ldots (\I-\Pi_{a_1^{\ell}}) = U_{a_m^{\ell}}\sqrt{\Theta_{a_m^{\ell}}}
\]
where $\Theta_{a_m^{\ell}}$ is some positive operator. Then, when Bob recovers the correct symbol $x=a_m^{\ell}$, the post measurement state is given by
\[
\frac{1}{\Tr\left[\Theta_{a_m^{\ell}}\rho_{a_m^{\ell}}\right]}U_{a_m^{\ell}}\sqrt{\Theta_{a_m^{\ell}}}\rho_{a_m^{\ell}}\sqrt{\Theta_{a_m^{\ell}}}\left(U_{a_m^{\ell}}\right)^{\dagger}
\]
Bob then applies the unitary $U_{a_m^{\ell}}$ to this state to get the state
\[
\frac{1}{\Tr\left[\Theta_{a_m^{\ell}}\rho_{a_m^{\ell}}\right]}\sqrt{\Theta_{a_m^{\ell}}}\rho_{a_m^{\ell}}\sqrt{\Theta_{a_m^{\ell}}}
\]
Suppose that $a_m^{\ell}=x\in \textsc{supp}(P'_X)$. Then, by the Gentle Measurement Lemma, we have that
\[
\norm{\rho_x-\frac{1}{\Tr\left[\Theta_{x}\rho_{x}\right]}\sqrt{\Theta_{x}}\rho_{x}\sqrt{\Theta_{x}}}_1 \leq 2\sqrt{\Tr\left[(\I-\Theta_{x})\rho_x\right]}
\]
It is now easy to see that the following bounds hold
\begin{align*}
    &\sum_xP_X(x)\norm{\rho_x-\frac{1}{\Tr\left[\Theta_{x}\rho_{x}\right]}\sqrt{\Theta_{x}}\rho_{x}\sqrt{\Theta_{x}}}_1 \\
    \leq & \sum_xP'_X(x)\norm{\rho_x-\frac{1}{\Tr\left[\Theta_{x}\rho_{x}\right]}\sqrt{\Theta_{x}}\rho_{x}\sqrt{\Theta_{x}}}_1+2\eps \\
    \leq & 2\sum_x P'_X(x)\sqrt{\Tr\left[(\I-\Theta_{x})\rho_x\right]}+2\eps \\
    \leq & 2 \sqrt{\sum_xP'_X(x)\Tr\left[(\I-\Theta_{x})\rho_x\right]}+2\eps \\
    \intertext{which implies, by our previous computations that, the above expression can be upper bounded by}
    \leq & 2\sqrt{\sqrt{2\eps}+\eps}+2\eps \eqqcolon \eps'
\end{align*}
The protocol then is that Bob, after decoding with the above POVM elements, places the classical symbol in the system $\hat{X}$. Let $\sigma^{X\hat{X}B}$ be the state after the protocol ends. We set 
the marginal $\sigma^{\hat{X}B}$ to be some junk if the decoding failed. Note that $\sigma$ is always classical on the system $X$. Then,
\begin{align*}
    &\norm{\rho^{X\hat{X}B}-\sigma^{X\hat{X}B}}_1 \\
    =&\sum_x P_X(x)\norm{\ketbra{x}^{\hat{X}}\otimes \rho_x^B-\sigma^{\hat{X}B}}_1 \\
    =&\E_{P_X}\left[\norm{\ketbra{x}^{\hat{X}}\otimes \rho_x^B-\sigma^{\hat{X}B}}_1\right] \\
    =& \E_{P_X}\left[\norm{\ketbra{x}^{\hat{X}}\otimes \rho_x^B-\sigma^{\hat{X}B}}_1~|~\textup{correct decoding}\right]\cdot \Pr[\textup{correct decoding}] \\
    &+\E_{P_X}\left[\norm{\ketbra{x}^{\hat{X}}\otimes \rho_x^B-\sigma^{\hat{X}B}}_1~|~\textup{incorrect decoding}\right]\cdot \Pr[\textup{incorrect decoding}] \\
    \leq & \sum_xP_X(x) \norm{\rho_x-\frac{1}{\Tr\left[\Theta_{x}\rho_{x}\right]}\sqrt{\Theta_{x}}\rho_{x}\sqrt{\Theta_{x}}}_1+\eps' \\
    \leq & 2\eps'
\end{align*}
This concludes the proof of the lemma.
\end{proof}

\section{Centralised Multi Link Measurement Compression}\label{sec:MeasurementCompression}

In this section we prove our centralised multi-link measurement compression theorem, in the presence of quantum side information. We precisely define the problem below:
Suppose that Alice possesses register $A$ of a pure state
$\ket{\rho}^{ABR}$.
Let $\Lambda^{A \to A XY} := \{\Lambda_{x,y}^{A \to A}\}_{x,y}$ be
a POVM where for each classical outcome $(x,y)$, $\Lambda_{x,y}$ is a
genuine POVM element i.e. a
Hermitian operator on $A$ with eigenvalues between zero and one.
There exist two separate noiseless channels, called $X$ and $Y$ channels,
from Alice to Bob, and 
two independent public coin registers, called $X$ and $Y$ public coins,
between them. A noiseless channel together with its corresponding 
public coin is called a link.
During the protocol at most one of the links may be turned OFF by 
an adversary without Alice or Bob's knowledge. Suppose Alice were
to measure register $A$ of state $\ket{\rho}^{ABR}$ with the POVM 
$\Lambda^{A \to A X Y}$ obtaining classical outcome $(x,y)$. 
In the centralised measurement compression protocol, Alice
compresses the outcome pair and conveys the messages through the
corresponding noiseless channels with the help of the corresponding
public coins. Let $R_X$, $R_Y$ denote the rates of the
noiseless channels for $X$ and $Y$ links, and $C_X$, $C_Y$ the rates for
the public coins for $X$ and $Y$ links. 
We require Bob to be able to decode with the help of the
public coins and produce a state on $ABRXY$ with the following properties:
\begin{enumerate}
	
\item
If the $X$ link is ON and $Y$ link is OFF, the state at the end of the
protocol should be $\epsilon$-close in Schatten $\ell_1$-norm to
$
\sum\limits_{x\in \mathcal{X}}
\ketbra{x}^X \otimes ((\Lambda_x \otimes \I^{BR})(\rho))^{A B R},
$
where $\Lambda_x^{A \to A} := \sum_y \Lambda_{x,y}$;

\item
If the $Y$ link is ON and $X$ link is OFF, the state at the end of the
protocol should be $\epsilon$-close in Schatten $\ell_1$-norm to
$
\sum\limits_{y\in \mathcal{Y}}
\ketbra{y}^Y \otimes ((\Lambda_y \otimes \I^{BR})(\rho))^{A B R},
$
where $\Lambda_y^{A \to A} := \sum_x \Lambda_{x,y}$;

\item
If both links are ON, the state at the end of the
protocol should be $\epsilon$-close in Schatten $\ell_1$-norm to the
state
$
\sum\limits_{(x,y)\in \CalX \times \mathcal{Y}}
\ketbra{x,y}^{XY} \otimes ((\Lambda_{x,y} \otimes \I^{BR})(\rho))^{A B R}.
$

\end{enumerate}
Moreover, Alice's and Bob's strategies should be agnostic to
which links are operational i.e. their encoding and decoding strategies
should continue to work even if one link fails.

We will prove \cref{thm:CentralisedMeasurementCompression} and \cref{cor:iidrateRegion}. We do this in several steps:
\begin{enumerate}
    \item We will first prove the centralised multi link measurement compression theorem in the \emph{one shot setting} when Bob does \emph{does} not posses any qunatum side information. To that end we will first show that a corner point in the rate region claimed in \cref{thm:CentralisedMeasurementCompression} (in the case when there is no side information) is achievable. See \cref{prop:1}.
    \item Using the techniques of \cref{prop:1} we show that a rate region of the kind claimed in \cref{thm:CentralisedMeasurementCompression} (in the case when there is no side information) is achievable by using the rate splitting technique shown in \cite{ChakrabortyNemaSen}. See \cref{prop:2}.
    \item We then derive the rate region when quantum side information is present at the decoder by composing the measurement compression theorem  with our protocol to do classical message compression with quantum side information. This is  a highly non-trivial task and we do this in \cref{subsec:MCSlepianWolf}.
    \item Finally, we observe that the rate region claimed in \cref{cor:iidrateRegion} is easily obtained by using the Quantum Asymptotic Equipartition Property (\cref{subsec:QAEP}) for all the one shot quantities used in the proof of the one shot theorem.
\end{enumerate}
\subsection{One Shot Centralised Multi Link Measurement Without Side Information}
\begin{proposition}\label{prop:1}
Given the state $\rho^A$ and the POVM $\Lambda_{XY}\coloneqq \brak{\Lambda_{xy}^A}$, the following rate point is achievable for centralised multi link measurement compression:
\begin{align*}
     R_X +  C_X &> \phmax^{\eps}(X) \\
     R_Y +  C_Y &> \phmax^{\eps}(Y) \\
     R_X &> I_{\max}^{\eps^4}(X:R) + O\big(\log \frac{1}{\eps}\big) \\
     R_Y &> I_{\max}^{\eps^4}(Y:RX) + O\big(\log \frac{1}{\eps}\big)\\
\end{align*}
where all entropic quantities are computed with respect to the state
\[
\sum\limits_{x,y}\ketbra{x}^X\otimes \ketbra{y}^Y\otimes \sqrt{\rho}\Lambda_{xy}\sqrt{\rho}^R
\]
\end{proposition}
\begin{proof}
\noindent{\bf POVM Construction} \\ \vspace{1mm} \\
We are given the POVM
\[
\Lambda_{XY}\coloneqq \brak{\Lambda_{xy}^A}
\]
and the state $\rho^A$. Consider the true post measurement state
\[
\sum\limits_{x,y}\ketbra{x}^X\otimes \ketbra{y}^Y\otimes \sqrt{\rho}\Lambda_{xy}\sqrt{\rho}^R
\]
We define 
\[
P_{XY}(x,y)\coloneqq \Tr[\Lambda_{xy}\rho]
\]
Let $\log K_1$ and $\log K_2$ be the number of public coins available to Alice and Bob with respect to the $X$ and $Y$ channels respectively. Similarly, let $\log L_1$ and $\log L_2$ be the number of bits that Alice needs to send Bob along the two channels respectively. We will show that the region indicated in the statement of the proposition is achievable for centralised multi link measurement compression. Construct $\mathcal{C}_X\coloneqq \brak{x(k_1,\ell_1)}$ and $\mathcal{C}_Y\coloneqq \brak{y(k_2,\ell_2)}$ randomly and independently from the distributions $P_X$ and $P_Y$, where $k_1\in [K_1], \ell_1\in [L_1]$ and $k_2\in [K_2], \ell_2\in [L_2]$. Define
\[
\rho(k_1,k_2,\ell_1,\ell_2)\coloneqq \frac{\sqrt{\rho}\Lambda_{x(k_1,\ell_1),y(k_2,\ell_2)}\sqrt{\rho}}{P_{XY}(x(k_1,\ell_1),y(k_2,\ell_2))}
\]
From the choices of $L_1$ and $L_2$ and Corollary \ref{corol:MeasTransCoveringLemma} we have that
\[
\E\limits_{\mathcal{C}_X\times \mathcal{C}_Y}\sum\limits_{k_1,k_2}\frac{1}{K_1\cdot K_2}\norm{\frac{1}{L_1\cdot L_2}\sum\limits_{\ell_1,\ell_2}\frac{P_{XY}(x(k_1,\ell_1),y(k_2,\ell_2))}{P_{X}(x(k_1,\ell_1))\cdot P_Y(y(k_2,\ell_2))}\rho(k_1,k_2,\ell_1,\ell_2)-\rho}_1 \leq \eps \tag{close}\label{eq:close}
\]
\begin{definition}\label{nice}
We call a block $(k_1,k_2)$ `nice' if the following condition holds for that block, with respect to some fixed codebook $\mathcal{C}_X\times \mathcal{C}_Y$:
\[
\norm{\frac{1}{L_1\cdot L_2}\sum\limits_{\ell_1,\ell_2}\frac{P_{XY}(x(k_1,\ell_1),y(k_2,\ell_2))}{P_{X}(x(k_1,\ell_1))\cdot P_Y(y(k_2,\ell_2))}\rho(k_1,k_2,\ell_1,\ell_2)-\rho}_1 \leq \sqrt{\eps}
\]
\end{definition}\vspace{2mm} \noindent Let us fix a nice block $(k_1,k_2)$ with respect to some fixed codebook. To ease the notation define
\[
t(k_1,k_2,\ell_1,\ell_2)\coloneqq \frac{P_{XY}(x(k_1,\ell_1),y(k_2,\ell_2))}{P_{X}(x(k_1,\ell_1))P_Y(y(k_2,\ell_2)}
\]
We are now in a position to apply \cref{corol:MeasureTransOpIneq}. To do this we define $Q$ to be the uniform distribution on the set $[L_1]\times [L_2]$. Then let
\[
Q'(k_1,k_2,\ell_1,\ell_2) \coloneqq \frac{Q(\ell_1,\ell_2)\cdot t(k_1,k_2,\ell_1,\ell_2)}{\sum\limits_{\ell_1,\ell_2}Q(\ell_1,\ell_2)\cdot t(k_1,k_2,\ell_1,\ell_2)}
\]
Then, \cref{corol:MeasureTransOpIneq}  implies that there exists a subset $\good \subset [L_1]\times [L_2]$ such that
\[
\Pr_{Q'}[\good] > 1-O(\eps^{1/4})
\]
and for all $(\ell_1,\ell_2)\in \good$ there exist states $\rho'(k_1,k_2,\ell_1,\ell_2)$ such that
\begin{align*}
    &\norm{\rho'(k_1,k_2,\ell_1,\ell_2)-\rho(k_1,k_2,\ell_1,\ell_2)}_1 \leq O(\eps^{1/4}) \\
    \intertext{and}
    & \sum\limits_{(\ell_1,\ell_2)\in \good} \frac{1}{L_1\cdot L_2} \cdot t(k_1,k_2,\ell_1,\ell_2) \rho'(k_1,k_2,\ell_1,\ell_2) \leq (1+O(\eps^{1/4}))\rho
\end{align*}
For the fixed block indices $k_1$ and $k_2$ we define the POVM elements for $(\ell_1,\ell_2)\in \good$:
\[
\gam{\ell_1,\ell_2}\coloneqq \frac{1}{1+O(\eps^{1/4})} \frac{t(k_1,k_2,\ell_1,\ell_2)}{L_1\cdot L_2}  \rho^{-1/2}\rho'(k_1,k_2,\ell_1,\ell_2)\rho^{-1/2}
\]
\vspace{2mm}\\
\textbf{Some Important Observations :}
\begin{align*}
    \sum\limits_{(\ell_1,\ell_2)\notin \good} \frac{t(k_1,k_2,\ell_1,\ell_2)}{L_1\cdot L_2} =& \left(\sum\limits_{\ell_1,\ell_2}Q(\ell_1,\ell_2)\cdot t(k_1,k_2,\ell_1,\ell_2)\right) \cdot \sum\limits_{(\ell_1,\ell_2)\notin \good} Q'(k_1,k_2,\ell_1,\ell_2) \\
\leq & (1+O(\eps^{1/4}))\cdot O(\eps^{1/4})    \\
=& O(\eps^{1/4})
\end{align*}
Then, we define the POVM 
\[
\Gamma(k_1,k_2)\coloneqq \brak{\gam{\ell_1,\ell_2}~|~(\ell_1,\ell_2)\in \good} \bigcup \brak{\gam{0}}
\]
where 
\[
\gam{0}\coloneqq \I_{\textup{supp}(\rho)   }-\sum\limits_{(\ell_1,\ell_2)\in \good} \gam{\ell_1,\ell_2}
\]
\begin{claim}
\[
\Tr[\gam{0}\rho]\leq O(\eps^{1/4})
\]
\end{claim}
\begin{proof}

\begin{align*}
    \Tr[\gam{0}\rho] =& \Tr[\rho-\sum\limits_{(\ell_1,\ell_2)\in \good}\frac{1}{1+O(\eps^{1/4})} \frac{t(k_1,k_2,\ell_1,\ell_2)}{L_1\cdot L_2}  \rho'(k_1,k_2,\ell_1,\ell_2)] \\
    =& 1- \frac{1}{1+O(\eps^{1/4})}\sum\limits_{(\ell_1,\ell_2)\in \good}\frac{t(k_1,k_2,\ell_1,\ell_2)}{L_1\cdot L_2}
\end{align*}
Recall that
\begin{align*}
    \sum\limits_{(\ell_1,\ell_2)}\frac{t(k_1,k_2,\ell_1,\ell_2)}{L_1\cdot L_2} \geq & 1-\eps \\
   \implies  \sum\limits_{(\ell_1,\ell_2)\in \good}\frac{t(k_1,k_2,\ell_1,\ell_2)}{L_1\cdot L_2} \geq & 1-\eps - \sum\limits_{(\ell_1,\ell_2)\notin \good}\frac{t(k_1,k_2,\ell_1,\ell_2)}{L_1\cdot L_2} \\
   \implies \sum\limits_{(\ell_1,\ell_2)\in \good}\frac{t(k_1,k_2,\ell_1,\ell_2)}{L_1\cdot L_2}  \geq & 1-\eps-O(\eps^{1/4})
\end{align*}
Therefore,
\begin{align*}
    \Tr[\gam{0}\rho] \leq 1-\frac{1-O(\eps^{1/4})}{1+O(\eps^{1/4})}\leq O(\eps^{1/4})
\end{align*}

\end{proof}

\vspace{2mm} 

\noindent{\bf Closeness of the Post Measurement States}\\ \vspace{2mm} \\
Alice will use the compressed POVMs only for those blocks which are nice. For all other blocks she aborts the protocol. Notice that the compressed POVMs that we designed output a classical index $(\ell_1,\ell_2)$. There exist functions $f$ and $g$ such that
\[
\begin{aligned}
f(k_1,\ell_1) \coloneqq & x(k_1,\ell_1) \\
g(k_2,\ell_2) \coloneqq & y(k_2,\ell_2)
\end{aligned}
\]
The error expression is then given  by
\begin{align*}
\begin{aligned}
    &\E\limits_{\mathcal{C}_X\times \mathcal{C}_Y}\left\lVert\sum_{x,y}\ketbra{x}^X\otimes \ketbra{y}^Y\otimes  \sqrt{\rho}\Lambda_{x,y}\sqrt{\rho}^R\right.\\ &\left.-\sum\limits_{\substack{(k_1,k_2)~\textup{ nice } \\
    (\ell_1,\ell_2)\in \good~\cup \brak{0}}}\frac{1}{K_1\cdot K_2}\ketbra{x(k_1,\ell_1)}\otimes \ketbra{y(k_2,\ell_2)}\otimes \sqrt{\rho}\gam{\ell_1,\ell_2}\sqrt{\rho}\right\rVert_1
    \end{aligned} \tag{post-measurement}\label{eq:postmeasure}
\end{align*}
\begin{definition}
We define random variable:
\begin{align*}
    T(\mathcal{C}_X,\mathcal{C}_Y)\coloneqq &\left\lVert\sum_{x,y}\ketbra{x}^X\otimes \ketbra{y}^Y\otimes  \sqrt{\rho}\Lambda_{x,y}\sqrt{\rho}^R\right.\\ &\left.-\sum\limits_{\substack{(k_1,k_2)~\textup{ nice } \\
    (\ell_1,\ell_2)\in \good~\cup \brak{0}}}\frac{1}{K_1\cdot K_2}\ketbra{x(k_1,\ell_1)}\otimes \ketbra{y(k_2,\ell_2)}\otimes \sqrt{\rho}\gam{\ell_1,\ell_2}\sqrt{\rho}\right\rVert_1
\end{align*}
and the event:
\[
E\coloneqq \left\{\# \textup{ of nice blocks }\geq (1-\eps^{1/4})\cdot K_1\cdot K_2\right\}
\]
\end{definition}
\noindent Consider the following claim which is easy to prove:
\begin{claim}
\[
\Pr[E]\geq 1-\eps^{1/4}
\]
\end{claim}
The error expression can then be analysed as:
\[
\begin{aligned}
\E[T] =& \E[T\cdot \mathrm{1}_{E}]+\E[T\cdot \mathrm{1}_{E^c}] \\
\leq & \E[T\cdot \mathrm{1}_E]+2\cdot \Pr[E^c] \\
\leq & \E[T\cdot \mathrm{1}_E]+2\cdot \eps^{1/4}
\end{aligned}
\]
The first inequality used the fact that $T$ is at most $2$, since it is the $1$-norm between two states. We will thus bound $T(\mathcal{C}_X,\mathcal{C}_Y)$ for only those codebooks for which the event $E$ holds.

We will massage the second term inside the $1$-norm expression above. First, we discard the $0$-th outcome which adds at most $O(\eps^{1/4})$. Next, notice that
\[
\sqrt{\rho}\gam{\ell_1,\ell_2}\sqrt{\rho} = \frac{1}{1+O(\eps^{1/4})} \frac{t(k_1,k_2,\ell_1,\ell_2)}{L_1\cdot L_2}  \rho'(k_1,k_2,\ell_1,\ell_2)
\]
We will apply the triangle inequality twice. In the first we replace all the $\rho'(k_1,k_2,\ell_1,\ell_2)$ states with $\rho(k_1,k_2,\ell_1,\ell_2)$. This gives rise to the term
\begin{align*}
    \frac{1}{1+O(\eps^{1/4})}\sum\limits_{\substack{(k_1,k_2) \textup{ is nice} \\
    (\ell_1,\ell_2)\in \good}}\frac{1}{K_1\cdot K_2} \frac{t(k_1,k_2,\ell_1,\ell_2)}{L_1\cdot L_2}\norm{ \rho'(k_1,k_2,\ell_1,\ell_2)-\rho(k_1,k_2,\ell_1,\ell_2)}_1
\end{align*}
Due to the conditions in \cref{corol:MeasureTransOpIneq}, this can be bounded above by
\[
\frac{1}{1+O(\eps^{1/4})}\sum\limits_{\substack{(k_1,k_2) \textup{ is nice} \\
    (\ell_1,\ell_2)\in \good}}\frac{1}{K_1\cdot K_2} \frac{t(k_1,k_2,\ell_1,\ell_2)}{L_1\cdot L_2} O(\eps^{1/4})
\]
We already know from \cref{eq:close} that for a nice block index $(k_1,k_2)$
\[
\sum\limits_{\ell_1,\ell_2}\frac{t(k_1,k_2,\ell_1,\ell_2)}{L_1\cdot L_2} \leq 1+\eps
\]
Therefore,
\[
\begin{aligned}
\frac{O(\eps^{1/4})}{1+O(\eps^{1/4})}\sum\limits_{\substack{(k_1,k_2) \textup{ is nice} \\
    (\ell_1,\ell_2)\in \good}}\frac{1}{K_1\cdot K_2} \frac{t(k_1,k_2,\ell_1,\ell_2)}{L_1\cdot L_2} \leq & \frac{O(\eps^{1/4})}{1+O(\eps^{1/4})} (1+\eps) \\
    \leq & O(\eps^{1/4})
\end{aligned}
\]
These steps have allowed us to massage the second term inside the norm in \cref{eq:postmeasure} into
\[
\begin{aligned}
\frac{1}{1+O(\eps^{1/4})}\sum\limits_{\substack{(k_1,k_2)\textup{ is nice} \\
    (\ell_1,\ell_2)\in \good}}&\frac{1}{K_1\cdot K_2}\frac{t(k_1,k_2,\ell_1,\ell_2)}{L_1\cdot L_2}\ketbra{x(k_1,\ell_1)} \\ 
    \otimes &\ketbra{y(k_2,\ell_2)}\otimes \rho(k_1,k_2,\ell_1,\ell_2)
    \end{aligned}
\]
We further massage this term by adding those terms in the sum above which correspond to those $(\ell_1,\ell_2)$ which are not in the set $\good$. This adds an extra
\[
\frac{1}{1+O(\eps^{1/4})}\sum\limits_{(k_1,k_2) \textup{is nice }}\frac{1}{K_1\cdot K_2}\sum\limits_{(\ell_1,\ell_2)\notin \good}\frac{t(k_1,k_2,\ell_1,\ell_2)}{L_1\cdot L_2} \leq O(\eps^{1/4})
\]
Collating the arguments above, we see that the true post measurement state in \cref{eq:postmeasure} can be replaced with the state
\[
\begin{aligned}
T_{\textsc{good}}\coloneqq\frac{1}{1+O(\eps^{1/4})}\sum\limits_{\substack{(k_1,k_2)\textup{ is nice} \\
    (\ell_1,\ell_2)}}&\frac{1}{K_1\cdot K_2}\frac{t(k_1,k_2,\ell_1,\ell_2)}{L_1\cdot L_2}\ketbra{x(k_1,\ell_1)} \\ 
    \otimes &\ketbra{y(k_2,\ell_2)}\otimes \rho(k_1,k_2,\ell_1,\ell_2)
    \end{aligned}
\]
Finally, we will add the terms corresponding to the blocks which are not nice, i.e,
\[
\begin{aligned}
T_{\textsc{bad}}\coloneqq \frac{1}{1+O(\eps^{1/4})}\sum\limits_{\substack{(k_1,k_2)\textup{ not nice }\\ (\ell_1,\ell_2)}}&\frac{1}{K_1\cdot K_2}\frac{t(k_1,k_2,\ell_1,\ell_2)}{L_1\cdot L_2}\ketbra{x(k_1,\ell_1)} \\ 
    \otimes &\ketbra{y(k_2,\ell_2)}\otimes \rho(k_1,k_2,\ell_1,\ell_2)
    \end{aligned}
\]
Recall that since we only consider those codebooks for which the event $E$ holds, there are at most $\eps^{1/4}\cdot K_1\cdot K_2$ blocks which are not nice. Consider the random variable
\begin{align}\label{tbad}
    \norm{T_{\textsc{bad}}}_1 = \frac{1}{1+O(\eps^{1/4})}\sum\limits_{\substack{(k_1,k_2)\textup{ not nice }\\ (\ell_1,\ell_2)}}\frac{1}{K_1\cdot K_2}\frac{t(k_1,k_2,\ell_1,\ell_2)}{L_1\cdot L_2} \tag{bad\_blocks}
\end{align}
\vspace{2mm} {\bf Bounding $\norm{T_{\textsc{bad}}}_1$} \\
\noindent From the classical measure transformed covering lemma we know that
\begin{align*}
\begin{aligned}
    &\E\limits_{\mathcal{C}_X\times \mathcal{C}_Y}\left\lVert\sum_{x,y}\ketbra{x}^X\otimes \ketbra{y}^Y\otimes  \sqrt{\rho}\Lambda_{x,y}\sqrt{\rho}^R\right.\\ &\left.-\sum\limits_{\substack{(k_1,k_2) \\
    (\ell_1,\ell_2)}}\frac{t(k_1,k_2,\ell_1,\ell_2)}{K_1L_1\cdot K_2L_2}\ketbra{x(k_1,\ell_1)}\otimes \ketbra{y(k_2,\ell_2)}\otimes \rho(k_1,k_2,\ell_1,\ell_2)\right\rVert_1 \\
    \leq & ~ \eps
    \end{aligned}
\end{align*}
This implies that
\begin{align*}
\begin{aligned}
    &\E\limits_{\mathcal{C}_X\times \mathcal{C}_Y}\left\lVert\sum_{x,y}\ketbra{x}^X\otimes \ketbra{y}^Y\otimes  \sqrt{\rho}\Lambda_{x,y}\sqrt{\rho}^R\right.\\ &\left.-\sum\limits_{\substack{(k_1,k_2) \\
    (\ell_1,\ell_2)}}\frac{t(k_1,k_2,\ell_1,\ell_2)}{K_1L_1\cdot K_2L_2}\ketbra{x(k_1,\ell_1)}\otimes \ketbra{y(k_2,\ell_2)}\otimes \rho(k_1,k_2,\ell_1,\ell_2)\right\rVert_1\cdot \mathrm{1}_E \\
    \leq & ~ \eps
    \end{aligned}
\end{align*}
Thus a random codebook satisfies the event $E$ and the condition
\begin{align*}
\begin{aligned}
    &\left\lVert\sum_{x,y}\ketbra{x}^X\otimes \ketbra{y}^Y\otimes  \sqrt{\rho}\Lambda_{x,y}\sqrt{\rho}^R\right.\\ &\left.-\sum\limits_{\substack{(k_1,k_2) \\
    (\ell_1,\ell_2)}}\frac{t(k_1,k_2,\ell_1,\ell_2)}{K_1L_1\cdot K_2L_2}\ketbra{x(k_1,\ell_1)}\otimes \ketbra{y(k_2,\ell_2)}\otimes \rho(k_1,k_2,\ell_1,\ell_2)\right\rVert_1\\
    \leq & ~ \sqrt{\eps}
    \end{aligned}
\end{align*}
with probability at least $1-\sqrt{\eps}$. Fix such a codebook. Then for this good codebook it is easy to see that
\begin{align*}
\begin{aligned}
    &\left\lVert\sum_{x,y}\ketbra{x}^X\otimes \ketbra{y}^Y\otimes  \sqrt{\rho}\Lambda_{x,y}\sqrt{\rho}^R\right.\\ &\left.-\frac{1}{1+O(\eps^{1/4})}\sum\limits_{\substack{(k_1,k_2) \\
    (\ell_1,\ell_2)}}\frac{t(k_1,k_2,\ell_1,\ell_2)}{K_1L_1\cdot K_2L_2}\ketbra{x(k_1,\ell_1)}\otimes \ketbra{y(k_2,\ell_2)}\otimes \rho(k_1,k_2,\ell_1,\ell_2)\right\rVert_1\\
    \leq & ~ O(\eps^{1/4})
    \end{aligned}
\end{align*}
We can write the second term in the $1$-norm above as a sum $T_{\textsc{good}}$ and $T_{\textsc{bad}}$. Then using the monotonicity of the $1$-norm under trace, we see that
\begin{align*}
    \left\lvert1-\Tr[T_{\textsc{good}}]-\Tr[T_\textsc{bad}]\right\rvert_1 \leq O(\eps^{1/4})
\end{align*}
From our previous analysis, we know that 
\begin{align*}
    &\left\lvert\Tr[T_{\textsc{good}}] -\Tr\left(\sum\limits_{\substack{(k_1,k_2)~\textup{ nice } \\
    (\ell_1,\ell_2)\in \good~\cup \brak{0}}}\frac{1}{K_1\cdot K_2}\ketbra{x(k_1,\ell_1)}\otimes \ketbra{y(k_2,\ell_2)}\otimes \sqrt{\rho}\gam{\ell_1,\ell_2}\sqrt{\rho}\right)\right\rvert_1 \\
    \leq & O(\eps^{1/4})
\end{align*}
Recall that for our choice of codebook, there are at most $\eps^{1/4}$ fraction of bad blocks. This implies that the second term in the expression above is at least $1-\eps^{1/4}$. Therefore, we can conclude that
\[
\Tr[T_{\textsc{good}}]\geq 1-O(\eps^{1/4})
\]
Using this lower bound on the trace of $T_{\textsc{good}}$, it is easy to see that
\[
\Tr[T_{\textsc{bad}}]\leq O(\eps^{1/4})
\]
Collating all the arguments above, we see that we can replace the true post measurement state in \cref{eq:postmeasure} to $T_{\textsc{good}}+T_{\textsc{bad}}$ with an  additive error of at most $O(\eps^{1/4})$. Recall that we have already fixed a codebook which bounds the $1$-norm between the ideal post measurement state and $T_{\textsc{good}}+T_{\textsc{bad}}$ by $\sqrt{\eps}$. Thus, this implies that there exists with probability at least $1-\sqrt{\eps}$ a codebook which ensures that
\[
\E\limits_{\mathcal{C}_X\times \mathcal{C}_Y}[T(\mathcal{C}_X,\mathcal{C}_Y)]\leq O(\eps^{1/4})
\]
This concludes the proof.
\end{proof} 

We will now show how one can achieve a larger rate region in the one shot setting for centralised multi link measurement compression by using the technique of quantum rate splitting, as shown in \cite{ChakrabortyNemaSen}. 

\begin{proposition}\label{prop:2}
Suppose Alice is given the state $\rho^A$ and the POVM $\Lambda_XY$ as in \cref{prop:1}. Suppose that $P_XY$ is the distribution induced by the POVM. Let $(P^{\theta}_U,P^{\theta}_V,\max)$ be a split of the marginal $P_X$, as defined in \cite{ChakrabortyNemaSen,UrbankeRateSplitting} , for some parameter $\theta\in [0,1]$. Then one achievable rate region is obtained as the
union over a parameter $\theta \in [0,1]$ of the regions 
$S_\theta$ defined by:
\[
S_\theta~:~ 
\begin{aligned}
&R_X  ~~=~~  R_U + R_V \\
&\begin{aligned}R_U ~~ > ~~  &I_{\max}^{\eps}(U:R)  \\
	  &+ O(\log \epsilon^{-1})\end{aligned} \\
&\begin{aligned}R_Y  ~~>~~  &I_{\max}^{\eps}(Y:RU)  \\
	  &+ O(\log \epsilon^{-1})\end{aligned} \\
&\begin{aligned}R_V  ~~>~~  &I_{\max}^{\eps}(V:RUY)  \\
	  &+ O(\log \epsilon^{-1})\end{aligned} \\
&C_X  ~~=~~  C_U + C_V \\
C_U + &R_U  ~~>~~  \phmax^{\eps}(U) \\
C_Y + &R_Y  ~~>~~  \phmax^{\eps}(Y) \\
C_V + &R_V  ~~>~~  \phmax^{\eps}(V), \\
\end{aligned}
\]
where the entropic quantities are calculated for the control state
\[
\begin{aligned}
\sum\limits_{(u,v,y)\in \CalX \times \CalX \times \mathcal{Y}}
&p^\theta_U(u) p^\theta_V(v) p(y|u,v)
\ketbra{u,v,y}^{UVY} \\
&\otimes 
\frac{((\Lambda_{\max\{u,v\},y} \otimes \I^{R})(\rho))^{A R}}
     {\Tr[((\Lambda_{\max\{u,v\},y} \otimes \I^{R})(\rho))^{A R}]}.
     \end{aligned}
\]
The above state is 
obtained by {\em splitting} random variable $X$ into independent random 
variables $U$, $V$ in the state
$
\sum\limits_{(x,y)\in \CalX \times \mathcal{Y}}
\ketbra{x,y}^{XY} \otimes ((\Lambda_{x,y} \otimes \I^{R})(\rho))^{A R}
$
according to the parameter $\theta$. 
Another achievability
region is obtained by rate splitting $Y$ instead of $X$. The total
achievable region is the union of the two regions. The encoding
and decoding strategies are agnostic to which links are actually
functioning.
\end{proposition}
\begin{proof}
First consider the post measurement state
\[
\sum\limits_{(x,y)\in \CalX \times \mathcal{Y}}
\ketbra{x,y}^{XY} \otimes ((\Lambda_{x,y} \otimes \I^{R})(\rho))^{A R}
\]
This can be rewritten as:
\[
\sum\limits_{(x,y)\in \CalX \times \mathcal{Y}}P_X(x)P_{Y|X}(y|x)
\ketbra{x,y}^{XY} \otimes \frac{((\Lambda_{x,y} \otimes \I^{R})(\rho))^{A R}}{\Tr\left[((\Lambda_{x,y} \otimes \I^{R})(\rho))^{A R}\right]}
\]
where 
\[
P_{XY}(x,y)\coloneqq \Tr\left[((\Lambda_{x,y} \otimes \I^{R})(\rho))^{A R}\right]
\]
One can then split the distribution $P_X$ into the distributions $p^{\theta}_U(u)$ and $p^{\theta}_V(v)$ such that
\[
\max\left(U,V\right)\sim P_X
\]
where the  $p^{\theta}_U(u)$ and $p^{\theta}_V(v)$ are both defined on the classical alphabet $\mathcal{X}$. Refer to \cite{ChakrabortyNemaSen} for details about this splitting operation. This leads to the control state:
\[
\begin{aligned}
\sum\limits_{(u,v,y)\in \CalX \times \CalX \times \mathcal{Y}}
&p^\theta_U(u) p^\theta_V(v) p(y|u,v)
\ketbra{u,v,y}^{UVY} \\
&\otimes 
\frac{((\Lambda_{\max\{u,v\},y} \otimes \I^{R})(\rho))^{A R}}
     {\Tr[((\Lambda_{\max\{u,v\},y} \otimes \I^{R})(\rho))^{A R}]}.
     \end{aligned}
\]
where
\[
p(y|u,v)\coloneqq P_{Y|X}(y|\max(u,v))
\]
Next, we construct the POVM elements as follows:
\begin{enumerate}
    \item Alice picks samples $\brak{U(k_1,\ell_1)~|~k_1\in [2^{C_U}], \ell_1\in[2^{R_U}]}$ iid from $p^{\theta}_U$, $\brak{Y(k_2,\ell_2)~|~k_1\in [2^{C_Y}], \ell_1\in[2^{R_Y}]}$ iid from $P_Y$ and $\brak{V(k_3,\ell_3)~|~k_3\in [2^{C_V}], \ell_1\in[2^{R_V}]}$ iid from $p^{\theta}_V$.
    \item Define 
    \[
    \Lambda_{u,v,y}\coloneqq \Lambda_{\max(u,v),y}
    \]
\end{enumerate}
The rest of the proof is similar to the proof of \cref{prop:1}, where in this case we have to use the sequential covering lemma for $3$ parties. Using the techniques in that proof, it is not hard to see that the rate region $S_{\theta}$ is achievable. This concludes the proof. This concludes the proof.
\end{proof}
\subsection{Measurement Compression with Side Information}\label{subsec:MCSlepianWolf}

In this section we will show how to compose the centralised measurement compression theorem with our protocol for classical data compression with quantum side information. We will first demonstrate our technique for the simpler case when the POVM only outputs one classical symbol $x$ i.e. the point to point case. 

\begin{proposition}\label{prop:3}
Given the shares quantum state $\rho^{AB}$, where the receiver Bob possesses the $B$ system, the following rates are achievable for one shot measurement compression with quantum side information:
\begin{align*}
R_X &> I_{\max}^{\eps}(X:RB) - I^{\epsilon_0/2}_H(X:B) 
	  + O(\log \epsilon^{-1})+1 \\
\intertext{and}
R_{X}+C_{X} &> \phmax^\eps(X) - I^{\epsilon_0/2}_H(X:B)+O(\log1/\eps)+1.
\end{align*}
where $\epsilon_0\coloneqq \eps^{1/10}$. Above, all entropic quantities are computed with 
respect to 
$
\sum\limits_{x\in \CalX}
\ketbra{x}^{X} \otimes ((\Lambda_{x} \otimes \I^{BR})(\rho))^{A B R}.
$
\end{proposition}
\begin{proof}
We will first construct a point to point measurement compression protocol which ignores that Bob has any side information. This follows as a corollary of \cref{prop:1}. Recall that this construction consists of a random codebook $\mathcal{C}_X$ which is  divided into $K$ blocks, where each block contains $L$ elements. Note that
\[
\begin{aligned}
&\log K+\log L > \phmax^{\eps}(X)-O(\log \eps) \\
&\log L > I_{\max}^{\eps}(X:RB)-O(\log \eps)
\end{aligned}
\]
where the entropic quantities are computed with respect to the control state
\[
\sum\limits_{x\in \CalX}
\ketbra{x}^{X} \otimes ((\Lambda_{x} \otimes \I^{BR})(\rho))^{A B R}.
\]
Suppose that $\mathcal{C}_X$ obeys the condition that the fraction of nice blocks is at least $1-O(\eps^{1/4})$, where a `nice' block refers to a block which satisfies \cref{nice}. Suppose that for every nice block index $k$, let \good(k) be the set of indices $\ell$ which obey the conditions of \cref{lem:opineq}. Then, our protocol ensures that with respect to this codebook, the global post measurement state is
\[
\Tilde{\rho}^{KLRB}\coloneqq\frac{1}{\Tr[\Tilde{\rho}]}\sum_{\substack{k\textup{ is a nice block}\\ \ell \in \good(k)}}\frac{1}{L\cdot K}\ketbra{k}^K\otimes \ketbra{\ell}^{L}\otimes \rho_{k,\ell}^{'RB}
\]
with the promise that for all nice indices $k$,
\[
\sum\limits_{\ell\in \good(k)}\frac{1}{L}\rho_{k,\ell}^{'RB}\leq (1+\eps^{1/4})\Tr_{A}((\Lambda \otimes \I^{BR})(\rho))^{A B R} 
\]

Recall that conditioned on the codebook, there exists a deterministic map $f : [K]\times [L]\to \mathcal{X}$ such that

\[
\norm{\rho^{XBR}-\Tilde{\rho}^{XBR}}\leq O(\eps^{1/4})
\]

Let $\Tilde{P}_X$ be the marginal of $\Tilde{\rho}^{XBR}$ on the system $X$. Then it is not hard to see that there exists a subset $\textsc{index}$ of $x$'s with probability at least $1-O(\eps^{1/8})$ under the distribution $P_X$ such that, for all $x\in \textsc{index}$

\[
\Tilde{P}_X(x) \leq (1+O(\eps^{1/8})) P_X(x)
\]

We now require that Alice and Bob work with a post-measurement state which has $\Tilde{P}_X$ as the marginal distribution on the system $X$. To do this, Alice simply discards those $(k,\ell)$ pairs which map to those $x$'s under $f$ which are not in $\textsc{index}$. We will call this post measurement state $\sigma^{KLB}$:
\[
\sigma^{KLRB}\coloneqq \frac{1}{\Tr[\sigma]}\sum\limits_{\substack{k \textup{ is nice}\\ \ell \in \good(k) \\ f(k,\ell)\in \textsc{index}} } \frac{1}{K\cdot L}\ketbra{k}^K\otimes \ketbra{\ell}^L\otimes \sigma^{'RB}_{k,\ell}
\]
where
\[
\sigma^{'RB}_{k,\ell}\coloneqq \rho^{'RB}_{k,\ell}
\]
and for all nice $k$,
\[\label{eq:op-ineq-1}
\sum\limits_{\substack{\ell\in \good(k) \\ f(k,\ell)\in \textsc{index}}} \frac{1}{L}\sigma^{'RB}_{k,\ell}\leq (1+\eps^{1/4})\Tr_{A}((\Lambda \otimes \I^{BR})(\rho))^{A B R}  \tag{op-ineq}
\]
Since $\Pr[\textsc{index}]\geq1-O(\eps^{1/8})$, it is not hard to see that
\[
\norm{\Tilde{\rho}^{KLRB}-\sigma^{KLRB}}_1 \leq O(\eps^{1/8})
\]
Note that the definition of $\sigma$ implies that, under the map $f$:
\[
\sigma^X \leq \frac{1}{1-O(\eps^{1/8})}\rho^X
\]
since the trace of $\sigma$ is at least $1-O(\eps^{1/8})$.

For the tools we develop in this section, we will not require the system $R$. Thus we will only  with the control state $\sigma^{KLB}$. Note that the operator inequality still holds since trace out preserves operator inequalities. \cref{eq:op-ineq-1} can now be written as:
\[
\sum\limits_{\substack{\ell\in \good(k) \\ f(k,\ell)\in \textsc{index}}}\frac{1}{L}\sigma_{k,\ell}^{'B}\leq (1+\eps^{1/4})\rho^B
\]
Now, consider the following claim:
\begin{claim}\label{claim:maincomposition}
Control state :
\[
\sigma^{KK'LB}\coloneqq\frac{1}{\Tr[\sigma]}\sum_{\substack{k\textup{ is a nice}\\ \ell \in \good(k) \\ f(k,\ell)\in \textsc{index}}}\frac{1}{L\cdot K}\ketbra{k}^K\otimes \ketbra{k}^{K'}\otimes \ketbra{\ell}^{L}\otimes \sigma_{k,\ell}^{'B}
\]
with the property that for every nice index $k$,
\[
\sum\limits_{\substack{\ell \in \good(k)\\f(k,\ell)\in \textsc{index}}}\frac{1}{L}\sigma_{k,\ell}^{'B} \leq (1+\eps^{1/4})\rho^B
\]
and under the map $f$
\[
\sigma^X \leq \frac{1}{1-O(\eps^{1/8})}\rho^X
\]
Then : 
\[
I_H^{\eps_0}(KL:BK')_{\sigma}\geq \log K+I_H^{\eps_0}(X:B)_{\rho^{XB}}-1
\]
where
\[
\eps_0\coloneqq \eps^{1/10}
\]
\end{claim}

\begin{proof}
Let $\Pi_{\textsc{OPT}}$ be the optimising operator in the definition of 
\[
D_{H}^{\eps_0}(\sigma^{KLB}~||~\sigma^{KL}\otimes \rho^B)
\]
Without loss of generality we can assume that
\[
\Pi_{\textsc{opt}}=\sum_{\substack{k\textup{ is a nice}\\ \ell \in \good(k) \\ f(k,\ell)\in \textsc{index}}}\ketbra{k}^K\otimes \ketbra{k}^{K'}\otimes \ketbra{\ell}^L\otimes \Pi_{k,\ell}^B
\]
For the purposed of brevity we will refer to the $(k,\ell)$ which obey the conditions in the definition of $\sigma$ as `fine'. Then
\begin{align*}
    \Pi_{\textsc{opt}}\sigma^{KL}\otimes \sigma^{K'B} &= c_0\cdot \Pi_{\textsc{opt}}\left(\sum_{(k,\ell)\textup{~fine~}}\frac{1}{L\cdot K}\ketbra{k}^K\otimes \ketbra{\ell}^L\right)\otimes \left(\sum_{k~\textup{fine~}}\frac{1}{K}\ketbra{k'}^{K'}\otimes \sigma_{k'}^{'B}\right) \\
    \intertext{where }
    \sigma_k^{'B}&\coloneqq \sum_{\ell\textup{ fine}}\frac{1}{L}\sigma_{k,\ell}^{'B} \\
    \intertext{and }
    c_0&\coloneqq \frac{1}{\left(\Tr[\sigma]\right)^2}\leq \frac{1}{\left(1-O(\eps^{1/8})\right)^2}  \leq \frac{1}{1-O(\eps^{1/8})}  \\
    \intertext{Then RHS is equal to}
    &= c_0\cdot \sum_{(k,\ell)\textup{~fine~}} \frac{1}{K^2\cdot L} \ketbra{k}^K\otimes\ketbra{k}^{K'}\otimes \ketbra{\ell}^L\otimes \Pi_{k,\ell}\sigma_k^B \tag{1} \\
    &\leq c_0\cdot(1+\eps^{1/4})\sum_{(k,\ell)\textup{~fine~}} \frac{1}{K^2\cdot L} \ketbra{k}^K\otimes\ketbra{k}^{K'}\otimes \ketbra{\ell}^L\otimes \Pi_{k,\ell}\rho^B 
\end{align*}
 Then taking trace on both sides and using the definition of $\Pi_{\textsc{OPT}}$ we get
\[
\begin{aligned}
2^{-I_{H}^{\eps_0}(KL:K'B)} \leq& c_0\cdot (1+\eps^{1/4})\Tr\left[\sum_{(k,\ell) \textup{ fine}} \frac{1}{K^2\cdot L} \ketbra{k}^K\otimes\ketbra{k}^{K'}\otimes \ketbra{\ell}^L\otimes \Pi_{k,\ell}\rho^B\right] \\
=& c_0\cdot (1+\eps^{1/4})\cdot 2^{-\log K}\cdot  2^{-D_H^{\eps_0}(\sigma^{KLB}~||~\sigma^{KL}\otimes \rho^B)} \\
\leq& c_0\cdot (1+\eps^{1/4})\cdot 2^{-\log K}\cdot 2^{-D_H^{\eps_0}(\sigma^{XB}~||~\sigma^X\otimes \rho^B)}
\end{aligned}
\]
where the last inequality is via the data processing inequality. \\
Next, suppose that $\Pi^{XB}$ is the optimising operator for 
\[
D_H^{\eps_0}(\sigma^{XB}~||~\sigma^X\otimes \rho^B)
\]
Also observe that
\[
\begin{aligned}
\Tr[\Pi^{XB}\sigma^{XB}] \geq & \Tr[\Pi^{XB}\rho^{XB}]-\norm{\sigma^{XB}-\rho^{XB}}_1 \\
\geq & 1-\eps_0-O(\eps^{1/8}) \\
\geq & 1-\frac{\eps_0}{2}
\end{aligned}
\]
where the last line is due to the fact that $\eps_0=\eps^{1/10}$. This implies that
\[
\begin{aligned}
2^{-I_{H}^{\eps_0}(KL:K'B)} \leq& c_0\cdot (1+\eps^{1/4}) \cdot 2^{-\log K}\cdot 2^{-D_H^{\eps_0/2}(\rho^{XB}~||~\sigma^X\otimes \rho^B)}
\end{aligned}
\]
Finally, since we know that 
\[
\sigma^X \leq \frac{1}{1-O(\eps^{1/8})}\rho^X
\]
one can see that
\[
\begin{aligned}
2^{-I_{H}^{\eps_0}(KL:K'B)} \leq& c_0\cdot (1+\eps^{1/4})\cdot \frac{1}{1-O(\eps^{1/8})} \cdot 2^{-\log K}\cdot 2^{-D_H^{\eps_0/2}(\rho^{XB}~||~\rho^X\otimes \rho^B)} \\
= & c_0\cdot (1+\eps^{1/4})\cdot \frac{1}{1-O(\eps^{1/8})} \cdot 2^{-\log K}\cdot 2^{-I_H^{\eps_0/2}(X:B)_{\rho^{XB}}}
\end{aligned}
\]
For small enough $\eps$, we can sue the following bound
\[
c_0\cdot (1+\eps^{1/4})\cdot \frac{1}{1-O(\eps^{1/8})} \leq 2
\]
Therefore, it is now clear that
\[
I_H^{\eps_0}(KL:K'B)_{\sigma}\geq \log K +I^{\eps_0/2}_H(X:B)_{\rho^{XB}}-1
\]
This concludes the proof.
 \end{proof}
 \noindent\textbf{Composing the Two Protocols :} \\
We will now use the claims proved above to show that the measurement compression theorem and the one-shot protocol for classical message compression with side information can be composed in the case when Bob has some side information. In that case, the actual post measurement state that we will work with is
\[
\sigma^{KLK'RB}\coloneqq \frac{1}{\Tr[\sigma]} \sum \frac{1}{K\cdot L} \ketbra{k}^K\otimes \ketbra{k}^{K'}\otimes \ketbra{\ell}^{L}\otimes \sigma^{'RB}_{k,\ell}
\]
where the summation is over the appropriate set of $(k,\ell)$ as defined before. We will treat the systems $K'$ and $B$ as side information available to Bob. One can now see that the arguments of \cref{claim:maincomposition} and \cref{thm:SlepianWolf} together imply that Alice has to send at most the following number of bits:
 \[
 \begin{aligned}
 &\phmax^{\eps_0}(KL)_{\sigma} - I_{H}^{\eps_0}(KL:K'R)_{\sigma} \\
 \leq & H^{\eps_0}_{\max}(KL)-\log K -I_{H}^{\eps_0/2}(X:B)_{\rho^{XB}}+1 \\
 \leq & \log K +\log L -\log K -I_{H}^{\eps_0/2}(X:B)_{\rho^{XB}}+1 \\
 = & \log L -I_{H}^{\eps_0/2}(X:B)_{\rho^{XB}}+1
 \end{aligned}
 \]
 From the measurement compression theorem, we know that
 \[
 \log L > I_{\max}^{\eps}(X:RB)-O(\log\eps)
 \]
 Thus, this implies that Alice needs to send at most
 \[
 I_{\max}^{\eps}(X:RB)-I_{H}^{\eps_0/2}(X:B)_{\rho^{XB}}-O(\log\eps)+1
 \]
 bits. This concludes the proof.
\end{proof}
\begin{remark}\label{remark:composition}
It is important to note that \cref{claim:maincomposition} holds even when the control state is a weighted sum of the following form:
\[
\sigma^{KK'LB}=\frac{1}{\Tr[\sigma]}\sum\frac{w(k,\ell)}{L\cdot K}\ketbra{k}^K\otimes \ketbra{k}^{K'}\otimes \ketbra{\ell}^{L}\otimes \sigma_{k,\ell}^{'B}
\]
This is precisely the case that we will face when we use \cref{claim:maincomposition} to prove \cref{thm:CentralisedMeasurementCompression}, due to the presence of the terms $t(k_1,k_2,\ell_1,\ell_2)$ (refer to the proof of \cref{prop:1}). However, it is easy to verify that even in this case the proof of \cref{claim:maincomposition} remains unchanged.
\end{remark}
We are finally ready to prove \cref{thm:CentralisedMeasurementCompression}.
\subsection{Proof of \cref{thm:CentralisedMeasurementCompression}}\label{subsec:finalproof}
\begin{proof}
The proof is not hard given the discussion in the previous sections. We will show that the region $S_{\theta}$, when $\theta=0$ is achievable. This argument can then be extended to other values of $\theta$ using the similar reasoning to what we used in proving \cref{prop:2}. To show that $S_0$ is achievable, we first invoke \cref{prop:1} to show that there exists a measurement compression scheme which achieves the rates given by $S_0$ in the case when there is no side information. To be precise, $S_0$ (in the case when the system $B$  is trivial)  is given by:
\begin{align*}
    R_X &> I_{\max}^{\eps}(X:RB) \\
    R_Y &> I_{\max}^{\eps}(Y:XRB)\\
    C_X+R_X &> \phmax^{\eps}(X) \\
    C_Y+R_Y &> \phmax^{\eps}(Y)
\end{align*}
where we have ignored the additive $O\big(\log \frac{1}{\eps}\big)$ factors. We will call the above region
\[
S_0^{\textup{unassisted}}
\]
Now when Bob possesses some quantum side information in the $B$ system, we can repeat the proof of \cref{prop:3} independently for the $X-$ and $Y-$ channels. This is possible because of two reasons:
\begin{enumerate}
    \item The encoding maps for the $X-$ and $Y-$ channels are independent by design.
    \item Even when both channels are ON, Bob can first decode the $X-$ channel and then decode the $Y-$ channel sequentially, since the first decoding does not perturb the state in the $B$ system too much. To be precise, this will lead to at most an additive error of some $g(\eps)$, where $g(\eps)$ is $\eps$ to some rational power. This is due to the Gentle Measurement Lemma and the proof of \cref{thm:SlepianWolf}.
\end{enumerate}
An important point to note that is that we cannot do any better in the sum rate, even though we could potentially get even more savings by repeating the proof of \cref{prop:3} in the case when both channels are ON. This is because we require the protocol to be agnostic to the actions of the adversary. For example, suppose Alice were to send a bit string along the $X-$ channel which was shorter than the one prescribed by our protocol. She would then have to compensate by sending more bits along the $Y-$ channel. However, the adversary could then simply turn OFF the $Y-$ channel, leading to a situation where it is not certain whether Bob can decode.

This is why, the rate region will look 
\[
S_0^{\textup{unassisted}}-(I_H^{\eps_0}(X:B),I_H^{\eps_0}(Y:B))
\]
The operation $S-(x_0,y_0)$ is defined as 
\[
S-(x_0,y_0)\coloneqq \brak{(a-x_0,b-y_0)~|~(a,b)\in S}
\]
For other values of $\theta$, we can now split the distribution $P_X$, as in the proof of \cref{prop:2}. Bob can then repeat his sequential decoding algorithm pretending as if there are three channels namely the $U-$ channel, the $Y-$ channel and the $V-$ channel. This shows that the rate region claimed in the statement of \cref{thm:CentralisedMeasurementCompression} is indeed achievable. This concludes the proof.
\end{proof}
\section{Asymptotic IID Analysis}\label{sec:iid}
In this section we will prove \cref{cor:iidrateRegion}:
\begin{proof}
As in the proof of \cref{thm:CentralisedMeasurementCompression}, we first show that all points in the region
\[
S_0^{\textup{assisted}}=S_0^{\textup{unassisted}}-(I_H^{\eps_0}(X:B),I_H^{\eps_0}(Y:B))
\]
is achievable when only $1$ copy of the state $\rho^{AB}$ is available. Now, we use the quantum asymptotic equipartition property for each of the one shot quantities in this expression to show that, when the state available for measurement is $\rho^{AB\otimes n}$, with $n\to \infty$. Thus, after normalising by $n$ and taking the limit, the region $S_0^{\textup{assisted}}$ is equivalent to:
\[
\begin{aligned}
R_X &> I(X:RB)-I(X:B) \\
R_Y &> I(Y:XRB)-I(Y:B) \\
C_X+R_X &> H(X)-I(X:B) \\
C_Y+R_Y &> H(Y)-I(Y:B)
\end{aligned}
\]
One can similarly prove that the following region, which we call $S_1^{\textup{assisted}}$, corresponding to $\theta=1$ is also achievable:
\[
\begin{aligned}
R_X &> I(X:YRB)-I(X:B) \\
R_Y &> I(Y:RB)-I(Y:B) \\
C_X+R_X &> H(X)-I(X:B) \\
C_Y+R_Y &> H(Y)-I(Y:B)
\end{aligned}
\]
One can then use a time sharing argument to show that the full achievable region is as follows:
\[
\begin{aligned}
R_X &> I(X:RB)-I(X:B) \\
R_Y &> I(Y:RB)-I(Y:B) \\
R_X+R_Y &> I(XY:RB)+I(X:Y) \\
&~~-I(X:B)-I(Y:B) \\
C_X+R_X &> H(X)-I(X:B) \\
C_Y+R_Y &> H(Y)-I(Y:B)
\end{aligned}
\]
This concludes the proof.
\end{proof}
\bibliography{ref}
\bibliographystyle{plain}

\appendix

\section*{Appendix}
\section{Proof of Lemma \ref{lem:SmoothMaxInfoEquiv}:}\label{sec:appendixA}
\begin{proof}
        Let $\sigma^{AB}\in\mathcal{B}^{\eps}(\rho^{AB})$. Let $k\coloneqq D_{\max}(\sigma^{AB}~||~\sigma^A\otimes \sigma^B)$. Then, by definition:
        \[
        \sigma^{AB}\leq 2^k \sigma^A\otimes \sigma^B.
        \]
        Define the operator:
        \[
        G^A\coloneqq \sigma^A+(\rho^A-\sigma^A)_+.
        \]
        It is not hard to see that $G^A\geq \sigma^A$ and $G^A\geq \rho^A$. Now, using $G^A$ we further define the operator:
        \[
        L^A\coloneqq \left(G^A\right)^{-1/2}\left(\left(G^A\right)^{1/2}\rho^A \left(G^A\right)^{1/2}\right)^{1/2}\left(G^A\right)^{-1/2}.
        \]
        This operator has the following useful properties which we will leverage throughout the proof:
        \begin{enumerate}
            \item $L^A$ is positive semi-definite by construction, since $G^A$ and $\rho^A$ are positive semi-definite.
            \item It holds that:
             \begin{align*}
            L^A \sigma^A L^{A} &\leq L^A G^A L^A \\
            &= \rho^A.
        \end{align*}
            \item $L^A \leq I$. To see this. consider that:
        \begin{align*}
            \left(G^A\right)^{1/2}\rho^A \left(G^A\right)^{1/2} &\leq \left(G^A\right)^{1/2}\left(G^A\right) \left(G^A\right)^{1/2}\\
            &= \left(G^A\right)^2.
        \end{align*}
        Using the definition of $L^A$ then implies the claim.
        \item $\Tr\left[\left(L^A\right)^2\sigma^A\right]\geq 1-\eps$. To see this, note that:
        \begin{align*}
            \Tr\left[L^A\sigma^AL^A\right] &\overset{(a)}{=} 1-\Tr\left[L^A\left(\rho^A-\sigma^A\right)_+L^A\right]\\
            &\overset{(b)}{\geq} 1-\Tr\left[\left(\rho^A-\sigma^A\right)_+\right]  \\
            &\overset{(c)}{\geq}1-\eps.
        \end{align*}
        In Step $(a)$ we have the used the definition of $G^A$ and the fact that $L^AG^AL^A= \rho^A$. In Step $(b)$ we have used the fact that $0\leq L^A\leq I$ and subsequently $\Tr\left[L^A\tau^A L^A\right]\leq \Tr\left[\tau^A\right]$ for any positive semi-definite $\tau^A$. In Step $(c)$ we have used the equivalence between the purified distance and the $1$-norm and the definition of the $1$-norm.
        \end{enumerate}
        Then, one can similarly define an operator $L^B$ with the same corresponding properties. Therefore:
        \[
        \wtt{\sigma}^{AB}\coloneqq \left(L^A\otimes L^B\right)\sigma^{AB}\left(L^A\otimes L^B\right) \leq 2^k \rho^A\otimes \rho^B.
        \]
        Note that:
        \begin{align*}
            I^{AB}-\left(L^A\right)^2\otimes \left(L^B\right)^2 &= I^{A}\otimes \left(I^B-\left(L^B\right)^2\right)+\left(I^A-\left(L^A\right)^2\right)\otimes \left(L^B\right)^2\\
            &\leq I^{A}\otimes \left(I^B-\left(L^B\right)^2\right)+\left(I^A-\left(L^A\right)^2\right)\otimes I^B,
        \end{align*}
        where the last inequality is due to the fact that $L^B\leq I^B$. Therefore:
        \begin{align*}
            \Tr\left[\wtt{\sigma}^{AB}\right] &\leq \Tr\left[\left(I^B-\left(L^B\right)^2\right)\sigma^B\right]+\Tr\left[\left(I^A-\left(L^A\right)^2\right)\sigma^A\right] \\
            &\leq 2\eps.
        \end{align*}
        Since $L^A$ and $L^B$ are positive semi-definite and $\leq I$, by the Gentle Measurement Lemma it holds that:
        \[
\norm{\sigma^{AB}-\frac{\wtt{\sigma}^{AB}}{\Tr\left[\wtt{\sigma}\right]}}_1 \leq 2\sqrt{2\eps}< 4\sqrt{\eps}.
        \]
        Again using the equivalence between the purified distance and the $1$-norm, we see that:
        \[
        P\left(\sigma^{AB}, \wtt{\sigma}^{AB}/\Tr\left[\wtt{\sigma}\right]\right) \leq 2\eps^{1/4}.
        \]
        This implies that $\wtt{\sigma}^{AB}/\Tr\left[\wtt{\sigma}\right]$ is a viable candidate in the optimisation in the definition of $J^{2\eps^{1/4}}(A : B)_{\rho}$. Normalising the main operator inequality on both sides by $\Tr\left[\wtt{\sigma}\right]$ and observing that this trace is at least $1-4\sqrt{\eps}$ (by the Gentle Operator Lemma), and using the definition of $J^{2\eps^{1/4}}$, we see that the following inequality holds:
        \[
        J^{2\eps^{1/4}}(A : B)_{\rho}\leq k + \log \frac{1}{1-4\sqrt{\eps}}.
        \]
        Recall that this $k$ is the function of the choice of $\sigma^{AB}$. The above derivation therefore implies that $J^{2\eps^{1/4}}-\log \frac{1}{1-4\sqrt{\eps}}$ is a lower bound for all choices of $\sigma^{AB}$, and hence is at most the infimum $I^{\eps}(A : B)_{\rho}$. This implies that:
        \[
        J^{2\eps^{1/4}}(A : B)_{\rho}\leq I^{\eps}_{\max}(A : B)_{\rho}+\log\frac{1}{1-4\sqrt{\eps}}.
        \]
        This concludes the proof.
        \end{proof}

        \section{Proof of Fact \ref{fact:smoothedConvexSplitLemma}}\label{sec:appendixB}
        \begin{proof}
            It is not hard to see that setting the parameters in Lemma 2 of \cite{WildeWiretap}  appropriately ($\eps\gets 36\eps^2$ and $\eta \gets 2\eps$), and by one application of the triangle inequality, we get that for:
            \[
            \log K \geq \widetilde{I}^{4\eps}_{\max}(B : A)_{\rho}+2\log \frac{1}{2\eps},
            \]
            it holds that:
            \[
            \norm{\tau^{A_1\ldots A_KB}-\rho^B\Otimes \rho^{A_{[K]}}}_1 \leq 20\eps
            \]
            Now, by Fact \ref{fact:imaxepsimax}, setting $\gamma\gets 2\eps$, we know that:
            \[
            \widetilde{I}^{4\eps}_{\max}(B:A)_{\rho}\leq J_{\max}^{2\eps}(A:B)_{\rho}+\log \frac{3}{4\eps^2}.
            \]
            Again, using Lemma \ref{lem:SmoothMaxInfoEquiv} we know that:
            \[
            J^{2\eps}(A:B)_{\rho}\leq I^{\eps^4}_{\max}(A:B)_{\rho}+\log \frac{1}{1-4\eps^2}.
            \]
            Therefore, the upper bound on the $1$-norm holds as long as:
            \[
            \log K\geq I^{\eps^4}_{\max}(A:B)_{\rho}+O\big(\log \frac{1}{\eps}\big).
            \]
            This concludes the proof.
        \end{proof}
\end{document}